\documentclass[11pt,a4paper]{article}

\usepackage{xr-hyper}
\usepackage{hyperref}

\usepackage[a4paper, left=2.0cm, right=2.0cm, top=2.5cm, bottom=2.5cm]{geometry}
\usepackage{amsmath,amssymb}   
\usepackage{graphicx}          
\usepackage{dcolumn}           

\usepackage{xcolor}
\usepackage{authblk}

\usepackage[mathlines]{lineno} 
\modulolinenumbers[5]

\providecommand{\keywords}[1]{\textbf{Keywords: } #1}
\newcommand{\tcr}{\textcolor{black}} 
\newcommand{\tcb}{\textcolor{black}} 

\title{Heatomics}

\author[1,2,3]{F.~Ritort\thanks{Contact, ritort@ub.edu, \url{https://felixritortlab.com/}}}

\affil[1]{Small Biosystems Lab, Condensed Matter Physics Department, Universitat de Barcelona\\
Carrer de Martí i Franques 1, 08028 Barcelona, Spain}

\affil[2]{Institut de Nanociència i Nanotecnologia (IN2UB), Universitat de Barcelona, Spain}

\affil[3]{Reial Acadèmia de Ciències i Arts de Barcelona (RACAB)\\
La Rambla 115, 08002 Barcelona, Spain}
\date{\today}
\begin{document}
\maketitle
\begin{abstract}Living cells are energy- and information-processing systems that sustain a nonequilibrium steady state (NESS) by continuously consuming energy and dissipating heat, as required by the second law of thermodynamics. \tcr{The rate of heat dissipation is closely related to the entropy production rate $\sigma$}, a universal primal life signal and unique descriptor of the cellular state. Living matter dissipates ${\cal P}_{\rm life}\sim$1 Watt/kilogram (W/kg), a remarkably conserved value across scales, from molecular reactions to entire organisms. Surprisingly, this high power density is $10^4$ times larger than that of the Sun and comparable to the universe’s average, ${\cal P}_{\mathcal{U}}= c^2H_0\sim 1$ W/kg, where $c$ is the speed of light and $H_0$ is Hubble's constant, a striking coincidence \tcb{reminiscent of} Dirac's large number hypothesis. \tcr{We surmise that this large ${\cal P}_{\rm life}$ sets the bar for generating negentropy, which contributes negatively to the overall positive $\sigma$. Negentropy sustains biological organization, distinguishing animate from inanimate matter.} Here, I introduce \textit{heatomics}, the science of studying $\sigma$ at the cellular and molecular scales, and the Variance Sum Rule, an experimental–theoretical framework that extracts $\sigma$ from fluctuations of a dynamical probe combined with the equation of state for a NESS. The emerging field of heatomics aims to elucidate the fundamental principles governing heat power generation, optimization of energy resources, and negentropy in living systems.\end{abstract}



\section{\label{sec:introduction}Introduction}
Living organisms constantly absorb and dissipate energy in non-equilibrium steady states (NESS) to avoid equilibrium, which means death. In living organisms, the generation of mechanical and chemical work is necessary to accomplish specific tasks, which unavoidably results in energy dissipation as heat, the most degraded yet universal form of energy and an essential signature of life. All living beings produce heat to maintain their homeostatic state. As a higher form of energy, work is task-specific and regulated, in contrast to the lower form of energy represented by heat, which is produced everywhere and always. Therefore, heat production is an essential and unavoidable part of all processes in the cell \cite{nicholls2013bioenergetics,yang2021physical,lan2012energy,cossetto2025thermodynamic}

To remain alive, cells use external energy sources through a myriad of molecular reactions to perform work and dissipate heat. From a thermodynamic viewpoint, the unique feature of living matter is its ability to self-organize, maintaining its integrity and resilience by staying permanently away from equilibrium through energy expenditure and heat production. Respiration is a key regulatory process in cells, where the energy of combustion of organic matter is used to synthesize ATP, the energy currency of metabolism. In the process, heat is produced across all scales, from single-molecule reactions to cells and tissues. 

\tcr{The rate at which heat is dissipated to the environment underlies the entropy production rate $\sigma$, life's universal and most essential signature.} Its measurement provides clues to understanding how life works, from the simplest to the most evolved organisms. The importance of heat for life cannot be overstated. Heat propagates by diffusion in water, a transport process dictated by the second law of thermodynamics, without need for any particularly evolved machinery or specific signaling molecules, making it the most basic signaling agent.  In addition, cells transition between different NESS in response to environmental changes, each characterized by a different value of $\sigma$, making it the natural descriptor of the cell's specific NESS. \tcr{In what follows, we contend that heat power is the distinguishing feature and main contribution to $\sigma$.} \tcb{Throughout the paper, we will use heat power and $\sigma$ interchangeably in isothermal environments, both expressed in units of Watts (W), omitting the temperature factor $T$ between entropy and heat.}

Despite its importance, measuring $\sigma$ at the molecular and cellular levels remains challenging. While work is the product of generalized forces and displacements and therefore experimentally accessible at the nanoscale, e.g., with force spectroscopy tools, heat cannot be directly measured from the same experiments. Heat is that part of the energy transferred to the hidden and largely inaccessible degrees of freedom. The overall positive $\sigma$ can be decomposed into the sum of positive and negative contributions corresponding to the different degrees of freedom. While most degrees of freedom contribute positively to $\sigma$, others can contribute negatively to reduce $\sigma$. This latter part often known as $\textit{negentropy}$ appears to us as the distinctive feature of life.  \tcr{This paper presents heatomics, the science of measuring $\sigma$ at the cellular and molecular scales and understanding the role of negentropy for living matter.} I describe the fundamental role of heat for life while discussing a recent novel approach to derive $\sigma$ based on the measurement of nonequilibrium fluctuations of a physical probe in contact with and coupled to the NESS. \tcr{Known as the Variance Sum Rule (VSR), the approach is applicable from molecular and cell biophysics to condensed matter physics, and beyond.}

The paper is organized as follows. In Sec.\ref{sec:whyheat}, I discuss the role of degrees of freedom for life energetics, heat, and negentropy. Sections \ref{sec:free} and \ref{sec:heat} overview the basics of the energetics of chemical reactions, the relation between heat and entropy production, and the conservation of the heat power density in living matter across scales. Section \ref{sec:NESS} discusses the Variance Sum Rule for the entropy production rate $\sigma$ and highlights recent applications to red blood cells. Section \ref{sec:heatomics} explores heatomics, the science of measuring the heat power in living matter and debates its relation to nanothermometry.  Section \ref{sec:negentropy}  examines the concept of negentropy in the context of polymerization reactions and heat transfer processes. Section \ref{sec:moluni} analyzes the role of heat power and negentropy for life, from molecules to the entire universe. Finally, sections \ref{sec:Discussion} and \ref{sec:conclusions} summarize the main results and discuss future perspectives in heatomics. 
\section{\label{sec:whyheat}Why heat?}
\tcb{Heat is essential for three reasons.} \tcr{First, heat $Q$ and its power $\dot{Q}\equiv \sigma$ is the most degraded, yet universal form of energy.} While one can talk about different types of work depending on the kind of forces, e.g., mechanical, electromagnetic, chemical, etc., heat is just one kind: random motion. Second, heat is omnipresent; it flows, tending to spread everywhere and always, as dictated by the second law, becoming the primal life signal. Third, heat production $\sigma$ permits the generation of negentropy and information, the proxies of life.    

Irreversibility, a signature of living matter, is determined by heat flows rather than by the exerted or extracted work. \tcb{Work represents ordered motion that can be completely converted into heat, or disordered motion, whereas the reverse conversion cannot occur completely without additional changes.} Heat corresponds to the energy of the internal degrees of freedom (hereafter abbreviated as DOF), that we cannot see, whereas work is the energy delivered or received by an external agent capable of exerting work. At the nanoscale, measuring heat is more challenging than measuring work, for which there exist force spectroscopy tools. 

One might wonder whether heat production is more relevant than other signatures related to metabolic processes, e.g., ATP consumption. \tcb{Besides ATP, metabolic reactions in the cell involve other heat-producing carriers such as ATP relatives (UTP, GTP,...), phosphate compounds, thioesters (Acetyl-CoA), redox carriers (NADH,...) among others.} If measuring heat at the nanoscale is already challenging, it is no less than measuring ATP consumption at the single molecule or sub-cellular level. Albeit related, heat production is the universal signature encompassing the consumption of all metabolic fuels.

\begin{figure*}[t]
    \centering
    \includegraphics[width=1\textwidth]{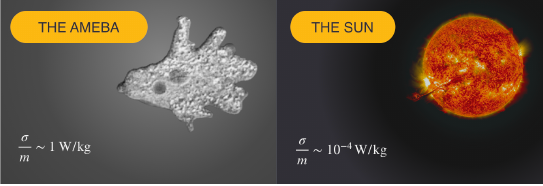}
    \caption{\textbf{Comparison of heat power densities between living matter and the Sun.} 
    }
    \label{fig:sun}
\end{figure*}

Central to understanding the heat power $\sigma$ in living matter is the concept of degree of freedom (DOF) denoted by $\omega$. A complex structure, such as a living cell, contains many DOF $\omega$ and a $\sigma$-density $\sigma_{\omega}$ describing each DOF's contribution to the total $\sigma$: $\sigma=\int d\omega \sigma_{\omega}$. The sum can be decomposed into two terms $\sigma_+$ and $\sigma_-$: $\sigma_+$ for the DOF ${\omega_+}$ that are exothermic, $\sigma_{\omega_+}>0$, and $\sigma_-$ arising from the DOF ${\omega_-}$ that are endothermic, $(-)\sigma_{\omega_-}<0$. The total $\sigma$ can be written as,
\begin{equation}\label{eq:sigmatotal}
    \sigma=\sigma_+-\sigma_-=\int_{\omega\in\omega_+}d\omega\sigma_{\omega}-\int_{\omega\in\omega_-}d\omega\sigma_{\omega}~ \ge 0.
\end{equation}
\tcr{We call the second term $\sigma_-$ the negentropy rate, or simply negentropy to abbreviate, the most essential signature of living matter.} The heat power $\sigma$ is an extensive quantity growing proportionally to the size of the system, so the value of $\sigma$ per unit mass is the relevant quantity. \tcb{The decomposition in \eqref{eq:sigmatotal} is unique once the relevant DOFs for a living system have been identified.} We argue below in Sec.\ref{sec:negentropy} that a large $\sigma$ per unit mass is a necessary condition for living matter. A large global $\sigma$ raises the bar for negentropy $\sigma_-$,  typically between 0.1-10\% of 
$\sigma$ while $\sigma_+\simeq \sigma$. In fact, it is quite unexpected to learn that, despite the large amount of energy released by nuclear reactions, the Sun has a heat power of $10^{-4}$ Watt/kilogram (W/kg), ten thousand times smaller than the 1 W/kg typical of living matter (Figure \ref{fig:sun}). For our Sun, this is the ratio between the total output power of $10^{26}$ W and the total Sun mass of $ 10^ {30} $ kg. Note that an average negentropy of 1$\%$ of the total $\sigma$ would amount to 0.01 W/kg for living matter, still a hundred times larger than the Sun's average $\sigma$-mass density. The role of negentropy for life is not of quantity, but of quality. In living matter, most energy-producing DOF are interacting and connected, leading to homeostasis and a universal highly conserved average heat power across scales \cite{Makarieva2008,Ballesteros2018}. A major question is how to measure negentropy and whether it is conserved across scales too. It is quite remarkable that the average heat power of the universe \tcr{estimated as $c^2H_0$ where c is the speed of light
and $H_0$ is the Hubble constant}, equals the average of $1$ W/kg, so characteristic of living matter. \tcr{This is discussed in Section \ref{sec:moluni}}.

\section{\label{sec:free}Free energy and all that}
\begin{figure*}[t]
    \centering
    \includegraphics[width=1\textwidth]{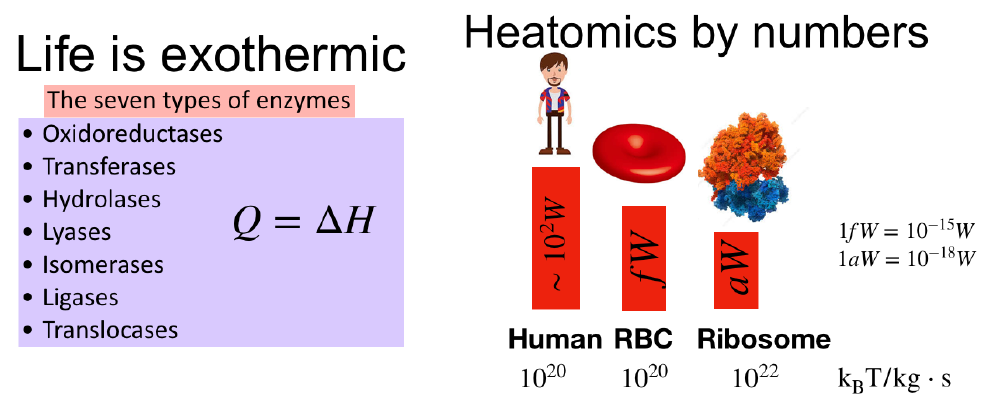}
    \caption{\textbf {\tcb{Enzyme} classification and heat power density in living matter.} (Left) Heat power is produced through metabolism by different types of enzymes. (Right) Heat power density is pretty conserved across scales and organisms.
    }
    \label{fig:enzymeshomeostasis}
\end{figure*}

 Metabolic reactions inside the cell are out of equilibrium, meaning that the free energy difference between reactants (r) and products (p) is non-zero, $\Delta G=G_{\rm p}-G_{\rm r}\ne 0$. Reactions with $\Delta G<0$ (exergonic) occur spontaneously according to the second law of thermodynamics. The free energy difference $\Delta G$ results from the balance between enthalpy and entropy, $\Delta G=\Delta H-T\Delta S$. For a dilute mixture kept at temperature $T$ and pressure $p$, the law of mass action holds and $\Delta G$ can be written as $\Delta G=k_BT\log({\cal Q}(T,p)/K_{\rm eq}(T,p))$ with ${\cal Q}$ the so-called reaction quotient (not to be confused with the heat $Q$) and $K_{\rm eq}$ the equilibrium constant. \tcr{In the above expression, the chemical potential of each species at concentration $c$ follows the law, $\mu(T,p)=\mu_0(T,p)+k_BT\log(c/c_0)$ with $c_0$ a reference concentration (typically 1 M). For a chemical reaction in thermodynamic equilibrium $K={\cal Q}$ so $\Delta G=0$.} Irreversible reactions inside the cell such as ATP hydrolysis are largely exothermic with ${\cal Q}/K_{\rm eq}\sim 10^9$ \tcb{and} an excess of reactants over products concentrations \cite{Atkins2023,Dill2010,Kondepudi2015}. A large fraction of the available $\Delta G$ is irreversibly released in the form of heat, $Q=-\Delta H$, with $\Delta H=H_p-H_r$ the enthalpy change for the reaction step $r\to p$. Here, the heat released to the environment is taken as positive, $Q>0$, corresponding to $\Delta H<0$, equal to the enthalpy decrease between reactants and products. \tcr{Throughout the paper, we use the sign convention dictated by the first law $\Delta U=W-Q$, where both work done on the system and heat released by the system to the environment raise the internal energy $U$.} For highly exothermic oxidative reactions, the entropy change $\Delta S=S_p-S_r$ is typically smaller than $Q/T$. An example is combustion of glucose into carbon dioxide and water, $\mathrm{C_6H_{12}O_6 + 6\,O_2 \rightarrow 6\,CO_2 + 6\,H_2O}$. In standard conditions, it generates a high amount of heat $Q=-\Delta H\simeq 700$ kcal/mol, roughly 97$\%$ of the available $\Delta G$ kcal/mol while $T\Delta S$ accounts only for the remaining 3$\%$. The combustion reaction in organic matter is performed at low rates through intermediate oxidation steps, with energy being stored in ATP, the energy currency of the cell. The energy stored in ATP is readily available through hydrolysis, $\mathrm{ATP + H_2O \rightarrow ADP + P_i}$. Under cellular conditions, ATP hydrolysis can reach $\Delta G\sim -15$ kcal/mol, depending on the conditions. In this case, $\Delta H$ and $T\Delta S$ are comparable; however, heat production $Q=-\Delta H$ still remains $60-70\%$ of the available $\Delta G$. This fact applies to most catalyzed highly exothermic reactions in the cell. According to the International Union of Biochemistry and Molecular Biology, enzymes are classified into seven families: oxidoreductases, transferases, hydrolases, lyases, isomerases, ligases and translocases (Figure \ref{fig:enzymeshomeostasis}, left panel). Except for isomerases where energies are negligible and translocases that are uphill ($\Delta G>0$), most of these enzymes catalyze downhill exergonic reactions ($\Delta G<0$) with values in the range \tcr{[-10:-3] kcal/mol often coupled to the ATP hydrolysis reaction \cite{Cox2025Lehninger}}. 

\section{\label{sec:heat}On the heat power $\sigma$}
When discussing metabolic reactions, it is important to distinguish between released heat $Q$ and entropy production $S_{\rm prod}$, which are closely related but not identical quantities. According to the second law, work $W$ is bounded by the free energy change, $W\ge \Delta G$. The difference $S_{\rm prod}=(W-\Delta G)/T$ defines the entropy production, a positive quantity for irreversible processes. When $W=\Delta G$, $S_{\rm prod}=0$ and the process is called reversible. For metabolic reactions where no work is involved, $W=0$, and $S_{\rm prod}=-\Delta G/T\ge 0$, \tcb{which} equals the available free energy divided by the temperature. 

In contrast, the released heat \tcb{directly equals the enthalpy change,} $Q=-\Delta H=-\Delta G-T\Delta S=TS_{\rm prod}-T\Delta S\ge -T\Delta S$. For open systems, $\Delta S=\Delta S_{\rm sys}+\Delta S_{\rm i-o}$ is the sum of the entropy change in the system, $\Delta S_{\rm sys}$, and the inflow and outflow entropy balance $\Delta S_{\rm i-o}(=\tcb{S_{\rm out}}-S_{\rm in})$ due to matter exchange with the environment. In an open-NESS, $\Delta S_{\rm sys}=0$ while $\Delta S_{\rm i-o}$ is non-zero, $Q\ge -T\Delta S_{\rm i-o}$ permitting both exothermic ($Q>0$) and endothermic ($Q<0$) reactions for $\Delta S_{\rm i-o}\ge 0$, e.g. in reactions where large molecules break up into many smaller ones. For closed-NESS systems, no matter exchange is allowed, so $\Delta S_{\rm sys}=\Delta S_{\rm i-o}=\Delta S=0$ and $Q=-\Delta H=-\Delta G=TS_{\rm prod}$ equals the entropy production times the temperature. Therefore, in a closed-NESS, the entropy production rate $\sigma=\dot{S}_{\rm prod}$ equals the heat power divided by the temperature, $\dot{Q}/T$, whereas for an open-NESS, $\dot{Q}=T(\sigma-\Delta \dot{S}_{\rm i-o})$ and heat power and entropy production rates differ. 

The main tenet of heatomics is that for most \tcb{metabolic processes}, the heat power $\dot{Q}$ remains the signature of $\sigma$, while $\Delta \dot{S}_{\rm i-o}$ typically remains a small fraction of the overall $\sigma$. Consequently, catabolism and oxidative reactions are highly exergonic and provide the main energy resources for life. 
Most relevant cellular processes consist of sequential and cyclic reactions of multiple steps. At each step, reactants flow in and products flow out (e.g. oxidation of NADH into NAD+, reduction of FAD into FADH$_2$, etc..), contributing positively and negatively to the overall $\Delta S_{\rm i-o}$, while some heat is released at all steps. 
An average human adult inhales and exhales air of different composition at a typical rate of about 6L/min in normal conditions, with $T\Delta \dot{S}_{\rm i-o}\sim 0.05$W/kg on average, i.e. less than 10$\%$ of the characteristic heat power 1W/kg, and $\sigma\simeq \dot{Q}/T$ within 90$\%$. 
\tcr{At the molecular scale, for instance during DNA polymerization, a monomer is added at every step while one ATP molecule is consumed at each step. The i-o mass balance contributes $T\Delta \dot{S}_{\rm i-o}\sim 1k_BT$, less than the typical $20k_BT$ of ATP hydrolysis. We emphasize that $T\Delta \dot{S}_{\rm i-o}$ is positive and adds to $\sigma_+$, in contrast to the opposite-sign negentropy that contributes to $\sigma_-$, cf. \eqref{eq:sigmatotal}.}


\subsection{\label{subsec:heatomics}Heatomics by numbers}
Molecular machines in the cell carry out specific tasks which are encoded in their architecture and inner workings. They do so either by rectifying thermal fluctuations through a Brownian ratchet mechanism or by catalyzing motion and work through a power stroke.  Feeding upon the energy and matter sources, molecular nanomachines, from translocases to polymerases and the ribosome, efficiently transduce the incoming energy and matter fluxes to generate useful work.  Embedded in their molecular architecture, information engines operate incessantly while producing heat. 

Energy dissipation and thermoregulation, the process by which living systems keep homeostatic conditions, holds across scales \cite{milo2015cell} (Fig. \ref{fig:enzymeshomeostasis}, right panel). \tcr{We remind that here $\sigma$ values are expressed in energy power units, such as Watts or $k_B T/s$, with $T$ the environmental temperature and $k_B$ the Boltzmann constant. Therefore, 1$k_BT/s\sim 4\times 10^{-21}$W at room temperature $T=25^{\circ}C\equiv 298K$.} The average $\sigma$ for the human body$\sim$1 W/kg is reflected down to the cellular and molecular scales, with the active fibroblasts and adipocytes dissipating on the scale picowatts (1pW=$10^{-12}$W), red blood cells (RBCs) dissipating $\sim$1 femtowatt (1fW=$10^{-15}$W), and the ribosome, the cellular machine in charge of protein synthesis that consumes the equivalent of four ATP molecules per added amino acid roughly every 0.1 s, dissipates $\sim$1 attowatt (1aW=$10^{-18}$W). The $\sigma$ values scale with system's size, giving $10^{20}\,k_B T\,/(\mathrm{kg}\cdot \mathrm{s})$ for the human and RBC cases, and $\sim 10^{22}k_BT/(\mathrm{kg}\cdot \mathrm{s})$ for the ribosome in consonance with its high metabolic activity. The above numbers are variable depending on the cell types and conditions. For instance, brown and beige adipocytes are natural heaters in mammals. They switch from a low to a high $\sigma$ state by activating the thermogenic pathway, increasing heat production to nW, which corresponds to hundreds of W/kg. At the level of tissues and organs, there is also variability; e.g., in humans, the heat power density of the brain is ten times larger, roughly $10$W/kg \cite{balasubramanian2021brain}. To be precise, the heat power density of living matter tends to vary but generally falls in the range $\sim$0.1-100 W/kg, with the rule of thumb of 1 W/kg, a highly conserved average power density across all life domains, from prokaryotes (bacteria) to eukaryotic cells (mammals, plants, fungi and protists). 
\section{\label{sec:NESS}Theoretical approaches to measure $\sigma$}
The experimental measurement of the entropy production rate $\sigma$ in NESS finds roots in the discipline of linear irreversible thermodynamics developed by Casimir, Onsager and the Dutch-Belgian school in the past century \cite{de2013non}, where $\sigma$ can be written in terms of sums of products of currents and forces of conserved physical quantities (e.g. mass, energy, momentum, charge), $\sigma=\sum_i J_iF_i$. Despite \tcb{extensive} theoretical work \cite{maes2003origin,falasco2025macroscopic}, experimental measurement of currents and forces at the nanoscale, and therefore of $\sigma$, has remained out of reach. In the last decade, alternative approaches to measure $\sigma$ have emerged in the framework of stochastic thermodynamics \cite{ritort2008nonequilibrium,seifert2012stochastic,ciliberto2017experiments,peliti2021stochastic}. The thermodynamic uncertainty relation (TUR) \cite{barato2015thermodynamic,horowitz2020thermodynamic,dechant2021improving}  sets a lower bound for the value of $\sigma$ in a NESS, $\sigma \ge \frac{2 k_B T \langle J \rangle^2}{t\, V_J(t)}$ which holds for all times, where $\langle J \rangle\equiv \langle J \rangle_t$ is the time-independent average current in the NESS and $V_J(t)$ stands for its variance measured over the time interval $t$,  $V_J(t) = \langle J_t^{2} \rangle - \langle J \rangle^{2}$. Despite its generality, the TUR bound is unpredictably loose, making $\sigma$ estimates unreliable \cite{rodriguez2025continuous}. In particular, flickering signals in a NESS such that $\langle J \rangle=0$ invariably give $\sigma\ge 0$, making the lower bound trivial. NESS irreversibility can be quantified through currents \cite{qian2004fluorescence,battle2016broken,bisker2017hierarchical,teza2020exact,li2019quantifying,manikandan2020inferring,roldan2021quantifying,dieball2022mathematical,pietzonka2024thermodynamic} yet accurate methods to measure the value of $\sigma$ are urgently needed. Fluctuation theorems and information-theoretical $\sigma$ estimates based on the Kullback-Leibler (KL) divergence rely on collecting a large number of trajectories and measuring $\sigma = \lim_{t \to \infty} \frac{k_B T}{t}\,\langle \log\!\left(\frac{P(\Gamma_t)}{P(\Gamma^{*}_t)}\right)\rangle$, i.e. the path-average of the logarithm of the probability ratio of forward $\Gamma_t$ and reverse $\Gamma^*_t$ paths of time duration $t$ \tcb{in the large-time limit} \cite{parrondo2009entropy,roldan2012entropy,skinner2021improved}.  
Besides the formidable task of experimentally monitoring systems with many degrees of freedom, the KL approach suffers from severe practical limitations. Observing time-reversed pairs of trajectories requires exceedingly long recurrence times, while the number of trajectories needed scales exponentially with $\sigma t$. These difficulties make the KL approach \tcb{impractical}. In fact, for typical heat powers in the range of femtowatts to picowatts, $\exp(\sigma t⁄k_B T)\approx 10^{100}$ trajectories are needed to estimate $\sigma\sim$fW for trajectories as short as 1 ms at room temperature ($1k_B T=4\times 10^{-21}$ J). TUR and KL-related approaches often yield values for $\sigma$ several orders of magnitude smaller than expected, e.g., on the order of pico-nanowatts for living cells. In comparison, a manageable problem arises when measuring equilibrium free-energy differences using the Jarzynski equality, e.g., the folding free energy $\Delta G$ of a biomolecule in pulling experiments, a question investigated in my lab over the years \cite{collin2005verification,junier2009recovery,alemany2012experimental,camunas2017experimental,rico2022molten,rissone2022stem}. \tcr{The exponential average of minus the work characteristic of the Jarzynski equality makes $\Delta G$ estimates strongly biased and methods are needed to reduce the bias for a finite number of experiments \cite{palassini2011improving,lee2026estimating}. In general,} the number of trajectories $N$ needed to estimate $\Delta G$ with $\sim k_B T$ accuracy grows exponentially with the dissipated work $N \sim \exp\left(\frac{W_d}{k_B T}\right)$, limiting the applicability to $\Delta G$ values of a few tens of $k_B T$ \cite{ritort2002two,gore2003bias}.

\section{\label{sec:VSR}The Variance Sum Rule (VSR)}
In the previous section, we discussed the challenge of using fluctuation relations to quantify irreversibility and to determine $\sigma$ \tcr{in conditions} where the total heat produced is much larger than $k_BT$, precluding efficient sampling of rare events under experimental conditions. Heat powers can always be measured using calorimeters; however, here we are talking about the nanoscale and tiny heat powers lower than nanowatts, the current benchmark for advanced nanocalorimetry \cite{lerchner2008nano,basta2018sensitive,hur2020sub,hong2020sub}. We will come back to this later in Section \ref{sec:heatomics}.

\begin{figure*}[t]
    \centering
    \includegraphics[width=0.6\textwidth]{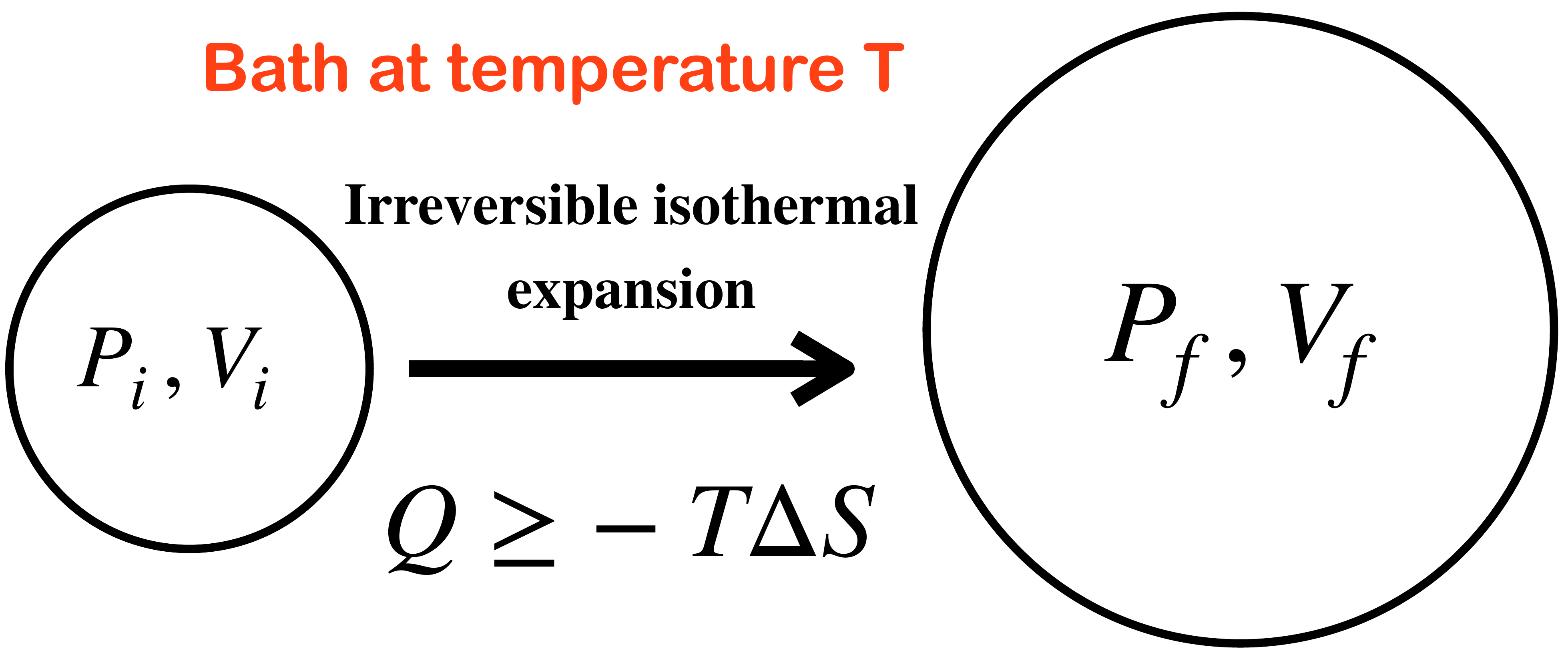}
    \caption{\textbf{Heat determination in irreversible processes.} Rather than using the Clausius inequality $Q\ge -T\Delta S$, which only gives a bound for the heat exchanged with the bath, thermodynamics employs energy conservation and a equation of state to determine the final state and the heat exchanged $Q$. 
    }
    \label{fig:energyconservation}
\end{figure*}

It is widely accepted that active fluctuations are the hallmark of nonequilibrium behavior; it is natural to seek alternative approaches that incorporate fluctuations while avoiding the problem of sampling rare events. The alternative approach to measuring $\sigma$ is known as the Variance Sum Rule (VSR)\cite{di2024variance,di2024variance2,roldan2024thermodynamic}. \tcb{The VSR reminds} of the approach of determining the heat exchanged in an irreversible transformation in classical thermodynamics. While heat is directly connected with entropy by the relation $dS=-\delta Q/T$ (here $\delta Q>0$ is the heat released by the system), for an irreversible isothermal process, the second law $dS\ge -\delta Q/T$ establishes that the change in entropy $\Delta S$ only results in a bound for the total heat $Q\ge -T\Delta S$. Therefore, the knowledge of $\Delta S$ obtained from the tabulated entropies for the initial and final states only gives a lower bound for $Q$. Rather than using the Clausius bound, it is preferable to use the first law of thermodynamics (energy conservation), $\Delta U=W-Q$, plus the system's equation of state (Fig.\ref{fig:energyconservation}). Unlike heat, work $W$ can always be expressed as the product of generalized forces and displacements and is experimentally measurable, also in the nanoscale.  \tcb{Moreover, the equation of state lets us determine $\Delta U$, from which $Q=W-\Delta U$ follows}. While the approach is model-dependent, it gives a good estimate for $Q$ provided the equation of state is sufficiently accurate. \tcb{The success of the Van der Waals equation in thermodynamics lies in the fact that, besides predicting phase transitions, it describes many liquid substances reasonably well with just two parameters}: the attractive intermolecular potential strength $a$ and the excluded volume $b$. \tcb{Similarly, the VSR sets a constraint on energy fluctuations due to active noise, which, together with an appropriate active model for the NESS equation of state, lets us predict $\sigma$.} As we will see below in Sec. \ref{subsec:model}, most active theoretical models employed to model the experiments are of the closed-NESS type, whereas metabolic reactions in living matter involve chemical flows and are open-NESS systems. \tcb{However, as discussed in Sec.\ref{sec:heat}, a major contribution to $\sigma$ is the heat power released to the environment, so we can neglect the entropy-flow part due to matter exchange to first order, making closed-NESS models suitable for this purpose.}  

\tcr{The VSR approach for determining $\sigma$ is based on monitoring the active fluctuations of a physicochemical probe that physically interacts with the system's DOF.} The probe can be mechanical, electromagnetic, or of another type, provided that its fluctuations are sensitive to the heat power generated inside the cell. \tcr{A requirement to apply the VSR is that the probe efficiently couples to the active DOF; otherwise, activity cannot be distinguished from Brownian passive noise.} For example, one might use a micron-sized bead captured in an optical trap in contact with the membrane of a living cell while monitoring its motion. \tcb{In this case, sensing the cellular activity requires that the stiffness of the optically trapped bead is lower than that of the cell membrane. Typical lipid bilayer} cell membranes have stiffness in the range of a few $pN/\mu m$.  Optical traps are ideal for sensing activity, as their stiffness can be tuned and reduced by decreasing the trapping laser power. In contrast, atomic force microscopy (AFM) cantilevers may not be appropriate because of their high stiffness, typically larger than $10^3pN/\mu m$, i.e., a thousand times larger than the membrane stiffness. However, one could also use electromagnetic fields via photon-emission detection to probe active fluctuations and measure $\sigma$ in living matter. A very informative experimental technique in cell biology is fluorescence lifetime imaging microscopy (FLIM), in which a fluorescently labeled molecule, e.g., a protein, is internalized into the cell \cite{noomnarm2009fluorescence,torrado2024fluorescence}. Upon illumination, the fluorophore is excited from its ground state to a higher energy state via light absorption and then re-emits a photon at a longer wavelength via spontaneous emission on the timescale of nanoseconds. \tcb{The energy levels and decay rates of the fluorescent protein are highly sensitive to internal cellular conditions, such as local pH and viscosity, which fluctuate due to the finite currents produced by the heat power $\sigma$ generated inside the cell. Such fluctuations directly affect the FLIM signal, at the levels of emission lifetime and intensity, which fluctuate on the timescale of cellular activity \cite{vallee2003molecular}}. This makes FLIM-flickering an ideal and essentially probe-free technique \cite{ma2024design}, as all fluorescent proteins respond to active electric fields via the pervasive atom-radiation dipolar interaction underlying all molecular forces, permitting us to determine $\sigma$. 

\subsection{\label{subsec:model}The VSR in a nutshell}
\begin{figure*}[t]
    \centering
    \includegraphics[width=0.8\textwidth]{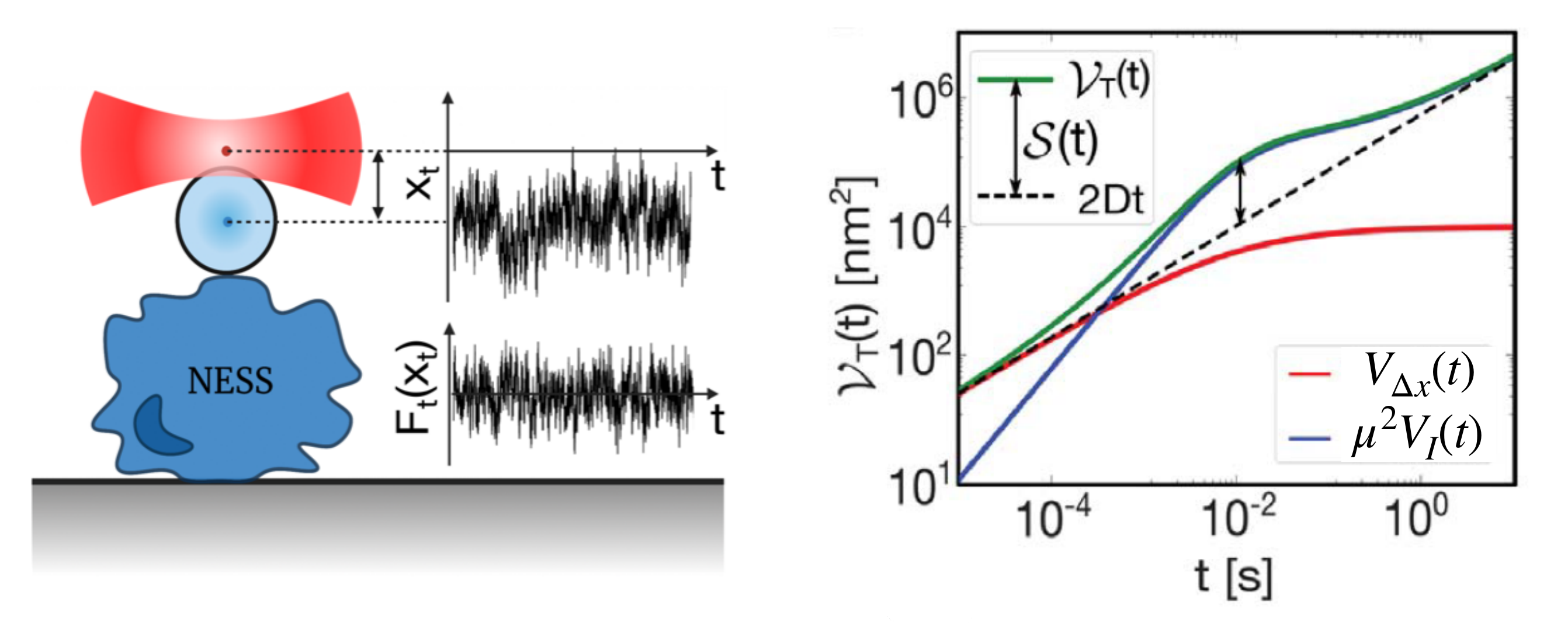}
    \caption{\textbf{The Variance Sum Rule (VSR)}(Left panel) A mechanical probe captured in an optical trap in contact with a NESS, e.g., a living cell, monitors the membrane flickering time-dependent position $x_t$, while $F_t$ stands for the net force acting on the bead. (Right) Illustration of the VSR showing the variances of the displacement ${\cal V}_{\Delta x}(t)$ and accumulated impulse ${\cal V}_{I}(t)$ during the time interval $t$. ${\cal V}_T$ denotes the sum of the two variances in \eqref{eq:VSR} while ${\cal S}(t)$ denotes the excess variance over the pure diffusion term $2Dt$. A NESS with positive $\sigma$ is characterized by a non-zero $S(t)$, cf. \eqref{eq:sigmaVSR}.
    }
    \label{fig:VSR}
\end{figure*}

The variance sum rule (VSR)\cite{di2024variance,di2024variance2} sets a constraint between the energy fluctuations and the dynamics of a flickering signal describing a few degrees of freedom in \tcb{a} NESS (Fig. \ref{fig:VSR}, left panel). The VSR considers the stochastic dynamics in a NESS of one or more degrees of freedom (DOF) in a noisy environment of passive and active forces. Probe dynamics sense the nonequilibrium system with which they are in contact, e.g., a bead tethered through polymeric linkers to a single enzyme catalyzing a chemical reaction or a bead in contact with the membrane of a single living cell. Bead dynamics are monitored and used to detect NESS activity to infer the heat power $\sigma$. In the simplest setting, the sensor is a small bead described by a single degree of freedom, whose motion is modeled by a Brownian particle of position $x_t$ and a time-dependent stochastic force $F_t$, measured \tcb{as a function of time $t$}. An overdamped Langevin equation describes \tcb{the dynamics},
\begin{equation}\label{eq:Langevin}
\dot{x}(t)=\mu F_t+\sqrt{2D}\eta_t
\end{equation}
where $\mu$ is the bead mobility, $D=k_BT\mu$ the diffusivity and $\eta_t$ a Gaussian white noise of zero average and correlation, $\langle\eta_t\eta_0\rangle=\delta(t)$ where $\langle...\rangle$ denotes the average over paths. Defined as $\sigma = \overline{{F}_t \circ v_t}$ with $v_t=\dot{x}_t$, the entropy production rate of a closed-NESS equals the NESS-averaged Stratonovich product \tcb{($\circ$)} of forces and velocities, $F_t$ and $v_t$, which are challenging quantities to measure at the nanoscale in the overdamped regime. In the framework of the VSR, fluctuations in the NESS are quantified by the variance of the particle's displacement $(\Delta x)$: $V_{\Delta x}(t)=\langle \Delta x_t^{2}\rangle$, where 
$\Delta x_t=x_t-x_0$ ($\langle\Delta x_t\rangle=0$) and the variance of the accumulated impulse $(I)$: $V_I(t)=\langle I_t^{2}\rangle-\langle I_t\rangle^{2}$ with 
$I_t=\int_{0}^{t}F_s\,ds$ the accumulated impulse between 0 and $t$.  Notice that both $\Delta x_t$ and $I_t$ are measured over a time interval of duration $t$. The VSR reads, 
\begin{equation}\label{eq:VSR}
V_T(t) \equiv V_{\Delta x}(t) + \mu^2 V_I(t) = 2Dt + S(t).
\end{equation}
The sum of the variances in the LHS of \eqref{eq:VSR} equals the thermal passive diffusion term $2Dt$ plus a term called excess variance $S(t)$ defined as
\begin{eqnarray}\label{eq:S}
    \begin{array}{rcl}
S(t)&=&2\mu\int_0^t du C_{xF}^a(u)   
        \\
        \\
        C_{xF}^a(u) 
        &=& \frac{1}{2}\bigl(C_{xF}(u)-C_{Fx}(u)\bigr) 
 \end{array}
\end{eqnarray}
and the one-time correlation function in the NESS, $C_{AB}(u)=\langle A(u)B(0)\rangle$. 

 The excess variance $S$ gives a direct estimate of $\sigma$ through its second time-derivative at $t=0$, 
\begin{equation}\label{eq:sigmaVSR}
 \sigma =\frac{1}{4\mu}\frac{d^2S(t)}{dt^2}\Bigr|_{t=0}\,\,\,.
\end{equation}

\begin{figure*}[t]
    \centering
    \includegraphics[width=1\textwidth]{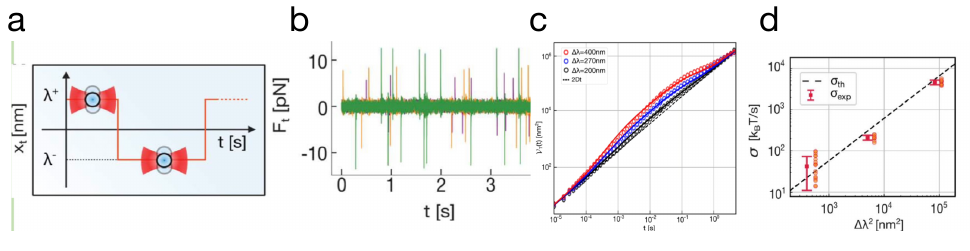}
    \caption{\textbf{The stochastic switching trap} (a) A bead captured in an optical trap jumping between two positions separated by $\Delta \lambda=\lambda_+-\lambda_-$. (b) Time-dependent net force $F_t$ for different values of $\Delta\lambda$ (different colors). The spikes represent the instantaneous jumps in trap position that increase the potential energy of the bead, subsequently dissipated as heat. (c) Test of the VSR for three different values of $\Delta \lambda$ (200nm, black), (270nm, blue), (400nm, red). The gentle curvature denotes the excess variance $S(t)$ in \eqref{eq:VSR}. (d) Heat power $\sigma$ versus $\Delta\lambda^2$ showing a linear dependence and spanning four orders of magnitude. Results taken from Ref.\cite{di2024variance}  
    }
    \label{fig:SST}
\end{figure*}
In equilibrium time-translational invariance imposes $C_{AB}(u)=C_{BA}(u)$ and $S(t)=\sigma=0$. 

The VSR \eqref{eq:VSR} informed by accurate experimental measurements of positions and forces permits the determination of $S(t)$ and the measurement of $\sigma$ from \eqref{eq:sigmaVSR}. \tcb{The VSR \eqref{eq:VSR} holds at all times $t$, typically from microseconds to seconds, imposing a dynamical constraint over six decades of time and enabling reliable derivation of $\sigma$ (Fig.\ref{fig:VSR}, right panel).} Figure \ref{fig:SST} exemplifies the VSR in an experimental toy model, a colloidal particle captured in an optical trap, whose center stochastically switches between two positions driving the particle to a NESS. The power injected by switching the optical trap's position is dissipated as heat to the environment. 

\begin{figure*}[t]
    \centering
    \includegraphics[width=0.8\textwidth]{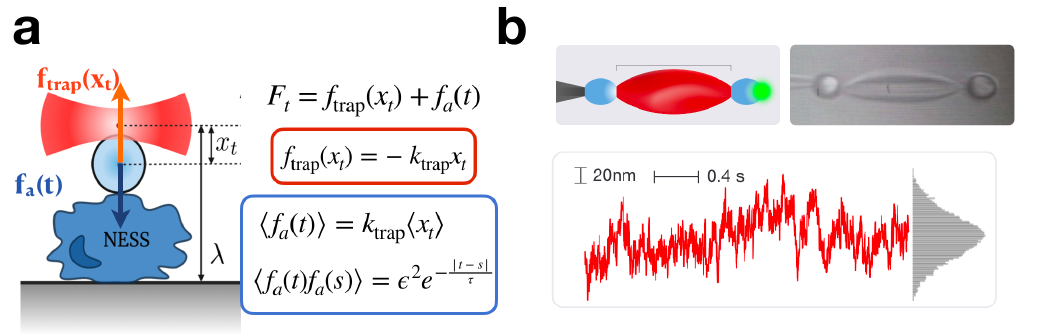}
    \caption{\textbf{The reduced-VSR.} (a) The net force $F_t$ acting on the bead is the sum of two opposite contributions\tcb{:} the force exerted by the optical trap and pointing upwards, $f_{\rm trap}(x_t)$ (red) and the active force exerted by the membrane of the living cell $f_a(t)$ (blue) pointing downwards. \tcb{In mechanical equilibrium, the average net force, $\langle F_t\rangle=0$, but the fluctuations expressed in the form of the variance $\langle F_t^2\rangle$ and $V_I(t)$ (cf. \eqref{eq:VSR}) remain undetermined, as only $f_{\rm trap}(x_t)$ can be measured.} The active force is modeled as a colored Gaussian noise (red and blue boxes). (b) Red blood cell mechanically stretched with optical tweezers \cite{gironella2024viscoelastic} and flickering signal of bead's position measured at a sampling rate of 100 kHz \cite{di2024variance}.
    }
    \label{fig:reducedVSR}
\end{figure*}

The main domain of applicability of the VSR is where the net force in \eqref{eq:Langevin} can be directly measured and $\sigma = \overline{{F}_t \circ v_t}$ estimated using the VSR. However, in most interesting situations this is not possible, and $\sigma$ remains inaccessible (Figure \ref{fig:reducedVSR}). In this case, \eqref{eq:VSR} can be recast in a reduced-VSR form in terms of the position $x_t$ alone, and the  measurement of the net force is not needed anymore,
\begin{equation}\label{eq:reducedVSR}
 V_{\Delta x}(t) + \mu^2 k^2\int_0^tdu\int_0^u dw V_{\Delta x}(w)= 2Dt + \tilde{S}(t)
\end{equation}
\tcb{where $k$ denotes the total stiffness of the probe and $\tilde{S}(t)$ is an effective excess variance that does not vanish in equilibrium and for which \eqref{eq:sigmaVSR} no longer holds.} The operation of eliminating DOFs that generate the active force in the NESS and result in \eqref{eq:reducedVSR} still requires a suitable active model that is solvable and which provides analytical expressions for $V_{\Delta x}(t)$, $\tilde{S}(t)$ and $\sigma$ in terms of the model parameters. The strategy to derive $\sigma$ consists in simultaneous fitting the model to \eqref{eq:reducedVSR}
and to the spectral density of the signal $x_t$, $\hat{C}(\omega)=|\langle\hat{x}_\omega|^2\rangle$ where $\hat{x}_\omega$ is the Fourier transform of $x_t$, to derive the  parameters of the model. The fitting parameters allow us to derive $\sigma$ from its analytical expression. The active model fitting \eqref{eq:reducedVSR} represents the equation of state of the NESS, while \eqref{eq:reducedVSR} sets a strong dynamical constraint over several decades of time that must contain the relevant active \tcb{timescales} to successfully derive $\sigma$. The feasibility of the reduced-VSR approach has been recently demonstrated for red blood cells (RBCs) \cite{di2024variance}; see Sec. \ref{subsec:RBC} and Figures \ref{fig:reducedVSR},\ref{fig:rigidity}, \ref{fig:RBC}. 
The VSR can also be extended to physical systems with hidden degrees of freedom, yielding a non-Markovian version, the so-called hidden-VSR \tcb{\cite{DiTerlizziInPrep}}, which enables the characterization of NESS in active viscoelastic media with multiple timescales, such as a cell's cytoplasm \cite{fernandez2006master,puig2007viscoelasticity,gironella2024viscoelastic,zhou2025viscoelastic}. The hidden-VSR measures the contributions by the hidden degrees of freedom to the total $\sigma$, a potentially suitable energy inference tool for complex systems, from machine learning to predictive AI. The current detection level of the VSR is 0.1 attowatts, or 100 zeptowatts (1 zW = $10^{-21}$ W; 1 $k_B T/s$), at room temperature (298K), which is appropriate for measuring $\sigma$ in molecular and cellular systems. \tcb{The VSR framework allows numerous extensions, particularly to discrete-state systems relevant to model probes based on electromagnetic fields, such as photon absorption and emission by a fluorescent molecule (e.g., the above-discussed technique of FLIM, fluorescence lifetime imaging microscopy).} Such systems can be modeled using a phenomenological three-state model, with excitation and decay kinetic rates described by a Jablonski diagram. 

\subsection{\label{subsec:RBC}Application to Red Blood Cells}
\begin{figure*}[t]
    \centering
    \includegraphics[width=1\textwidth]{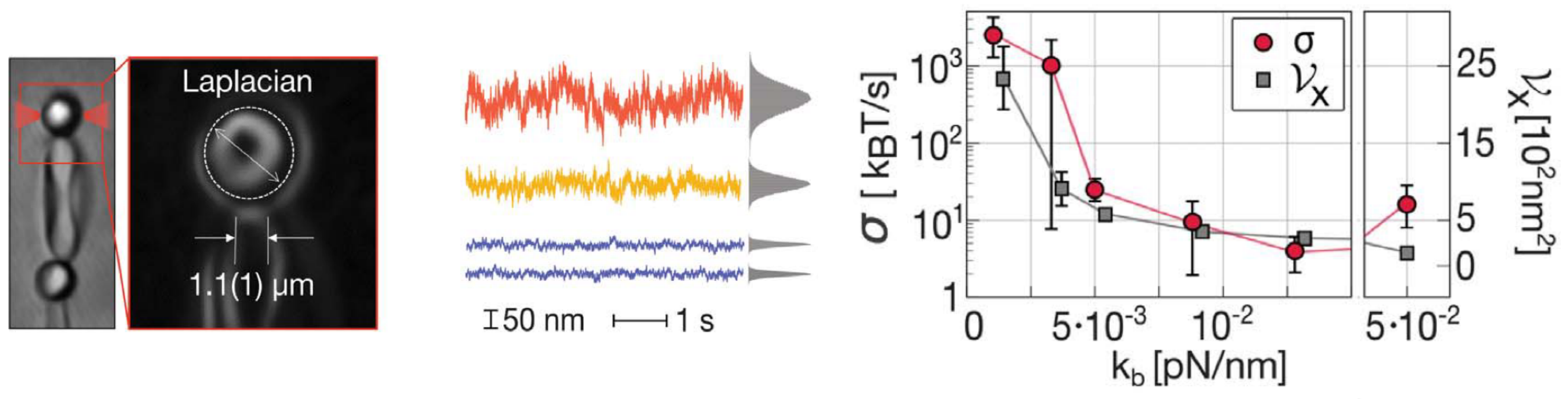}
    \caption{\textbf{Rigidity dependent probe sensitivity.} (Left and middle) Video image of a stretched RBC with the zoom showing the area of contact membrane-bead, $\sim 1\mu m^2$, and flickering traces with low trap stiffness (red and orange) and high trap stiffness (blue). (Right) Dependence of measured heat power $\sigma$ (left axis) and flickering bead position variance $V_x$ (right axis) with trap stiffness $k_b$. Only for $k_b\ll k_{RBC}$ with $k_{RBC}\sim 5pN/\mu m$ the RBC membrane rigidity \cite{gironella2024viscoelastic}) \tcb{can we} detect active fluctuations and reliably estimate $\sigma$. Results taken from Ref.\cite{di2024variance}.
    }
    \label{fig:rigidity}
\end{figure*}

\begin{figure*}[t]
    \centering
    \includegraphics[width=0.8\textwidth]{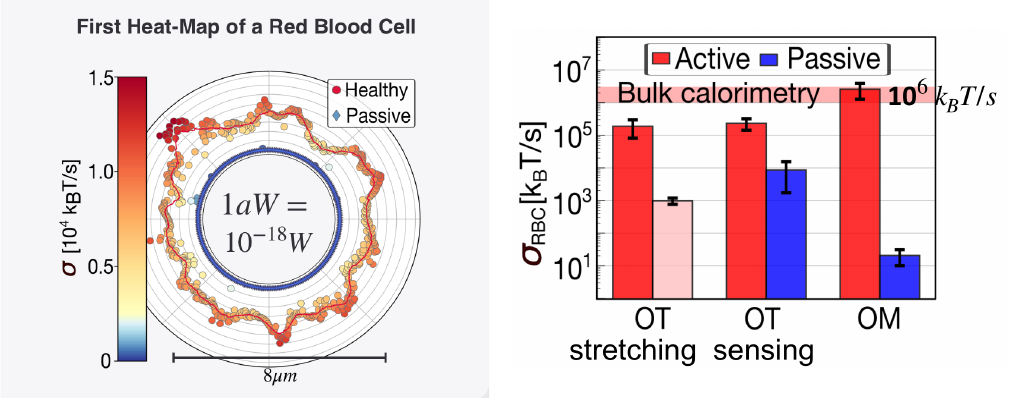}
    \caption{\textbf{Heat power map of a single RBC.} (Left) Heat power map over lateral regions of pixel size 50nmx50nm along the equatorial rim of an RBC. (Right) Comparison of heat power measured with bulk calorimetry (upper red bar) \cite{backman1992microcalorimetric}, \tcb{single-RBC measurements with optical tweezers (OT) by stretching and sensing,} and optical microscopy (OM). Active (red) versus passivated (blue) RBCs. \tcb{The light-red bar for OT-stretching measurements was obtained at high trap-power stiffness values} (Figure \ref{fig:rigidity}, right). Data from Ref.\cite{di2024variance}.
    }
    \label{fig:RBC}
\end{figure*}

\tcr{The VSR has been applied to derive $\sigma$ from flickering measurements of a mechanical probe in contact with the membrane of a red blood cell (RBC) (Fig.\ref{fig:reducedVSR}, right panel).} Probe dynamics senses the nonequilibrium system and is used to measure the heat power $\sigma$ released to the environment. The $\sigma$ value is robust and model-independent \tcb{as long} as it captures the timescale of the flickering activity. Moreover, we have shown that VSR's applicability requires the stiffness of the probe bead $k_b$ to be smaller than the RBC's rigidity $k_{RBC}$, \tcr{typically $\sim 5$ pN/$\mu$m}, otherwise if $k_b\gg k_{RBC}$ the fluctuations of the probe are dominated by passive noise, $\langle \Delta x_t^{2} \rangle = k_B T/(k_b+k_{RBC})\simeq k_B T/k_b$ masking RBC's activity, Fig. \ref{fig:rigidity}. 
With the VSR, it has been possible to measure the $\sigma$-map of single red blood cells using force spectroscopy and ultrafast imaging microscopy for the first time. The monitoring of the probe dynamics permitted to directly measure the $\sigma$-map of RBC \cite{di2024variance} (Fig.\ref{fig:RBC}, left panel) finding an average heat flux $j\sim 10^{4}\,k_B T/\mu\mathrm{m}^{2}\,\mathrm{s}$ and a total $\sigma \sim 2\,\mathrm{fW}\sim10^{6}\,k_B T/\mathrm{s}$ in the scale of femtowatts (fW) per single RBC in agreement with bulk studies \cite{backman1992microcalorimetric} \tcb{(Fig.\ref{fig:RBC}, right panel)}. \tcb{We stress that the VSR reports heat power, $\sigma$, rather than the intracellular temperature reported in nanothermometry studies.} The thermal conductivity and heat capacity of water, highest among liquids of comparable molecular weight, result in temperature variations smaller than $10^{-6}$ K across the cell. A crude estimation of the temperature variation $\Delta T$ generated by an RBC across its size $L\sim 5\mu m$, gives $\Delta T\sim \sigma\,/\kappa L \sim j\,L/\kappa \approx 10^{-9}\,\mathrm{K},$ with $\kappa = 0.6\,\mathrm{W\,m^{-1}\,K^{-1}}$ the thermal conductivity of water. The reported unphysical increase in intracellular temperatures, on the order of a few degrees in some cases, has drawn criticism from the standpoint of kinetic theory \cite{baffou2014critique}. In fact, temperature is an equilibrium concept, whereas kinetic temperature measurements require monitoring the average kinetic energy and the velocities of intracellular molecules, which are currently inaccessible. Overall, intracellular temperatures are just effective temperatures that depend on the probe and bulk calibration, proxies for $\sigma$ at best. In contrast, $\sigma$ is a physically grounded, probe-independent quantity (mechanical, fluorescent, etc.) that can be derived from flickering data, the VSR, and a suitable active model. The relation between intracellular temperatures and $\sigma$ is further elaborated in the next section.
\section{\label{sec:heatomics}On Heatomics}
\begin{figure*}[t]
    \centering
    \includegraphics[width=0.6\textwidth]{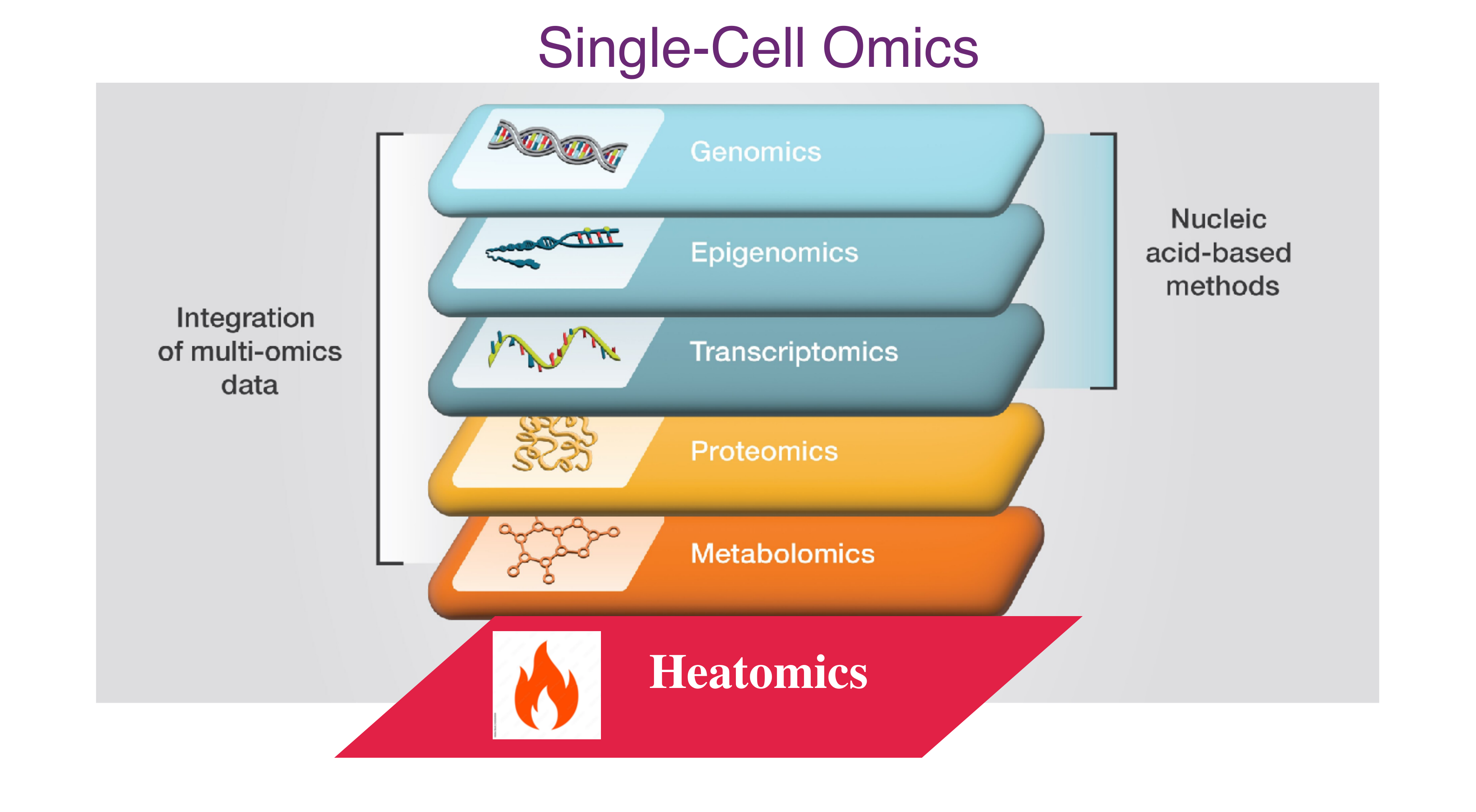}
    \caption{\textbf{Heatomics.} A few illustrative layers of the omics disciplines across biology. We posit that heatomics is a fundamental layer encompassing life processes.
    }
    \label{fig:heatomics}
\end{figure*}

Heatomics is the science of heat power in living matter. It introduces a new paradigm by integrating the science of thermodynamics in the form of heat, as a new layer of omics data into biophysics and cell biology, combining theoretical and experimental approaches for non-equilibrium systems. Omics layers in biology include biological components, such as genomics, metabolomics, and proteomics, and functional layers, such as interactomics, regulomics, and mechanobiomics (Fig.\ref{fig:heatomics}). Still, the most fundamental and integrative science, thermodynamics, is left out of such studies. \tcb{Energy is widely recognized as the universal and unifying quantity without which life is not possible.} Therefore, to the above \textit{omics} list it is natural to add thermodyn\underline{\bf{o}}mics (the underlined "\underline{\bf{o}}" is on purpose) or even better heatomics, which is shorter and emphasizes the specific role of heat power, the most degraded form of energy, in living matter. As Einstein once put it when talking about thermodynamics: \textit{It is the only physical theory of universal content, which I am convinced, that within the framework of applicability of its basic concepts will never be overthrown.} 

\tcb{Over the past decades, the nanothermometry field has attracted many scientists \cite{vetrone2010temperature,donner2012mapping,okabe2012intracellular,kucsko2013nanometre,ebrahimi2014nucleic,spicer2021harnessing,chuma2024implication}. Temperature measurements at the subcellular level have seemingly reported intracellular temperatures several degrees higher than the culture medium.} Methods for registering intracellular temperatures include polymeric nanoparticles, quantum dots, fluorescent probes, nanodiamonds, etc. \cite{thompson2018plug,zhou2020advances}. In the case of mitochondria, it has been reported that \tcb{their} temperature can even reach $50\,^{\circ}\mathrm{C}$ for thermostated cultures at $37\,^{\circ}\mathrm{C}$ \cite{chretien2018mitochondria}. \tcb{These local temperature deviations conflict with kinetic theory, which precludes such large temperature gradients over the micron-scale \cite{baffou2014critique}}. As we already observed in Sec.\ref{subsec:RBC}, the thermal conductivity of water $\kappa = 0.6\,\mathrm{W\,m^{-1}\,K^{-1}}$ ,  could not sustain such temperature deviations. For the largest $\sigma$ values, e.g., the nW of brown adipocytes with the activated thermogenic pathway, $\Delta T\sim \sigma\,/\kappa L \approx 10^{-3}\,\mathrm{K}$ falls in the mK range. 
\tcr{Only by close packing many cells in a sufficiently large volume $V$ can a significant increase in temperature be measured \cite{backman1992microcalorimetric}}. In fact, the total heat power $P$ is extensive with the number of cells $n$, $P=\sigma n$ with $n=V/v$ and $v$ the cell volume. Moreover, the typical heat diffusion length scale $L\sim V^{1/3}$ gives $\Delta T\sim P/\kappa L=n\sigma/\kappa L\sim\sigma V^{2/3}/v\kappa$. A typical brown adipocyte has a size of $\sim 30 \mu m$ and a volume $v\sim 10^4 \mu m^3$. Assuming a heat power of $\sigma=1$nW per single adipocyte, we would get an increase of one degree $\Delta T=1$ K for a cubic millimeter volume $V=(v \kappa\Delta T/\sigma)^{3/2}\sim 1$ mm$^{3}$ containing $n=V/v\sim 10^{5}$ fully packed adipocytes. \tcb{Brown adipose tissue can have dimensions of $L\sim 1$ cm, giving $\Delta T=10$ K, a strategy mammals use to maintain homeostatic conditions during hibernation.} Therefore, large $\Delta T$ values are found at the level of macroscopic thermogenic tissues but not at the single-cell level. In fact, the calorimetric measurements on RBCs, which we referred to in Sec.\ref{subsec:RBC} were made at the level of large volume samples ($\sim$ liter) in the heat-flow mode to keep the samples temperature constant. \tcb{Still, temperature measurements of densely packed RBCs under macroscopic conditions where $V=1$ liter, $L\sim 10$cm and $\sigma\sim 1-10$ fW,} a maximum temperature increase $\Delta T\approx 1-10$mK should be expected within the range of microcalorimetry measurements. 

\tcb{In summary, one cannot obtain an intracellular temperature increase of one degree at the single-cell level.}  What then makes the temperature appear so high in nanothermometry studies? \tcb{In thermodynamics, temperature is an equilibrium concept based on the zeroth law: all equilibrium systems in mutual contact with a bath share the common equilibrium bath temperature and have no heat flow between them.} The bath is an infinite reservoir of energy \tcb{whose} properties do not change in time, and that can be faithfully taken as equilibrated by construction. The equilibrium property is universally applicable to all systems whatever their size, from macroscopic systems to single atoms.  \tcb{In contrast,} nonequilibrium systems are characterized by net flows of conserved quantities such as energy. \tcb{Still, one can define} a local kinetic temperature in terms of the average kinetic energy of the molecules and the equipartition law, $k_BT=2E_{\rm kin}/g$, with $g$ the number of molecular DOF (3 for monoatomic, 5 for diatomic, and so on). \tcb{However, intracellular measurements based, e.g., on fluorescence imaging, do not measure the kinetic temperature but a proxy for the cell's internal activity based on fluorescence emission intensity, lifetime, or wavelength.} Calibration of these fluorescent nanothermometers is carried out outside the cell in equilibrium conditions. Nothing guarantees that these sensors will measure the local kinetic temperature \tcr{in a nonequilibrium environment}. First, because they do not locally report on the average kinetic energy of the surrounding water molecules that occupy about $80\%$ of the total cell volume; Second, because intracellular temperatures might be probe-dependent, i.e., a quantum dot may report a different value than a fluorescent protein or a fluorescent-quencher pair. Therefore, the intracellular temperatures are \emph{effective temperatures} at most. To the best of our knowledge, no test has been provided on equal values of intracellular temperatures obtained using different probes or light emission observables (wavelength, polarization anisotropy, intensity, lifetime, etc).  Besides, a living cell is a highly heterogeneous system with different structures and organelles at the sub-cellular level, from the cytoplasm to the nucleus, and from the mitochondria to the endoplasmic reticulum, peroxisomes, Golgi apparatus, etc. Measuring an intracellular temperature map of the cell and observing temperature variations across the cell would conflict with kinetic theory for the reasons given above. 

\begin{figure*}[t]
    \centering
    \includegraphics[width=1\textwidth]{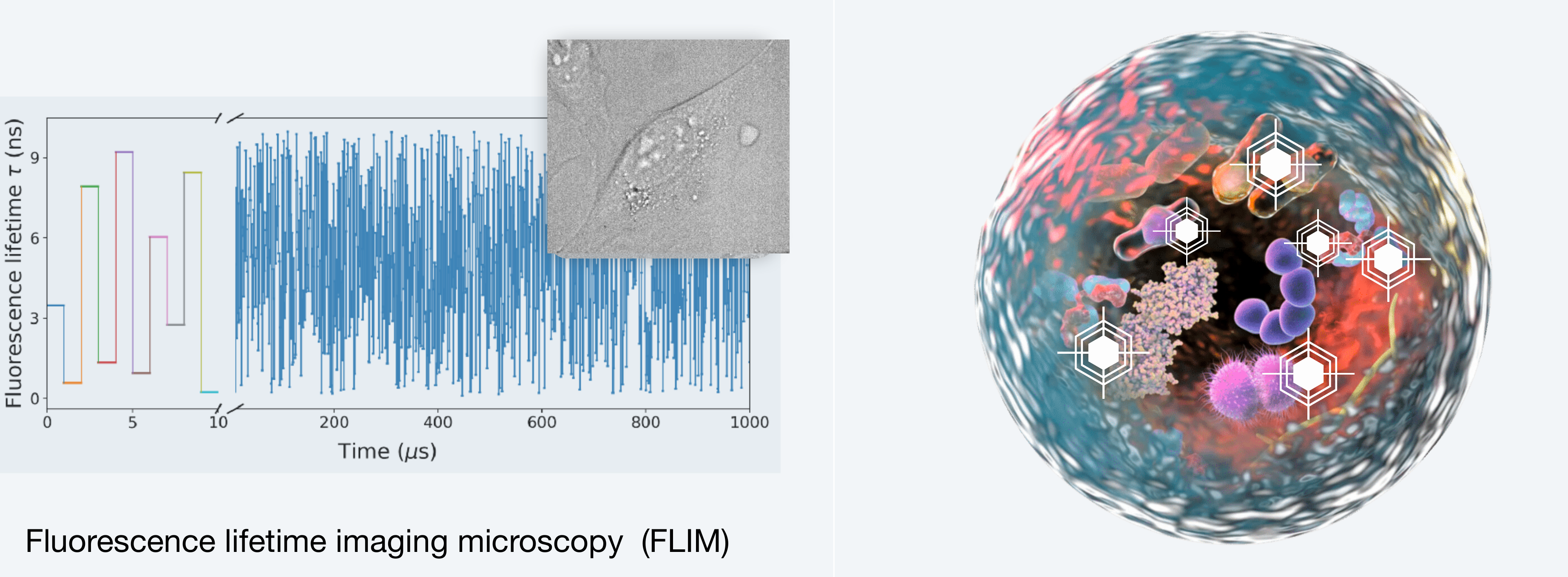}
    \caption{\textbf{$\sigma$-map of single cells.} (Left) Illustration of flickering lifetimes expected for FLIM (fluorescence lifetime imaging microscopy) measurements. Video image of a fibroblast illustrates how such measurement could be done with fluorescent proteins internalized at specific sub-cellular locations. (b) Illustration of \tcb{how the $\sigma$-map of a single cell could be obtained using} fluorescent probes internalized at different cellular locations.
    }
    \label{fig:heatmap}
\end{figure*}

\tcb{In contrast, the heat power is a well-defined quantity out of equilibrium. It is what micro- and nano-calorimeters measure, giving full meaning to $\sigma$.} The VSR approach takes a step further by combining measurements of the spectral density of the observable (denoted as $x_t$ in Sec.\ref{subsec:model}) and the VSR-based modeling to derive $\sigma$ with a resolution of attowatts (aW), $10^9$ times beyond the nanocalorimetry bar. In contrast to nanothermometry, the $\sigma$ values should be independent of the probe, and the same $\sigma$ should be obtained with a mechanical degree of freedom or a light-emission fluorescence signal. \tcb{We note that the same fluorescence techniques used for nanothermometry measurements may also apply to $\sigma$ measurements with the VSR.} For example, one could use fluorescence lifetime imaging microscopy (FLIM) in a confocal setup with organelle-targeted fluorescent proteins (e.g., to the nucleus, mitochondria, endoplasmic reticulum, peroxisomes), extending the VSR framework to FLIM-based flickering in living cells. Typical emission lifetimes of fluorescent proteins are in the nanosecond \tcb{scale. Moreover, measuring flickering lifetimes over $\mu$s timescales would provide a wealth of information about lifetime fluctuations, from which the heat power generated in the cell could be derived.} Measurements at different cell locations should provide the first $\sigma$-map of individual cells at the sub-cellular level (Fig.\ref{fig:heatmap}). Sub-cellular $\sigma$ measurements should be independent of the type of fluorescent probe, color, and illumination intensity, as in mechanical experiments where the measured $\sigma$ is independent of the probe used and setup, e.g. optical tweezers with stretching or sensing versus ultrafast microscopy (Fig.\ref{fig:RBC}, right panel). We emphasized in Sec.\ref{sec:VSR} that a good active model that reproduces the flickering spectrum and the VSR is essential for deriving $\sigma$. However, the value of $\sigma$ should not depend much on the active model implemented, provided it includes the relevant active timescales in the spectral density of the flickering signal, and its dynamics comply with the VSR equations over the NESS timescales. A further test of the VSR approach should demonstrate that both FLIM- and mechanical-based measurements give the same heat power $\sigma$ for single cells. The impact of these developments in biophysics and cell biology would be huge, as I discuss in Secs. \ref{sec:Discussion} and \ref{sec:conclusions}.  
\section{\label{sec:negentropy}On negentropy}
We have seen in Sec.\ref{sec:whyheat} that the total entropy production rate $\sigma$ in a NESS can be expressed as the sum of $\sigma(\omega)$ over the different degrees of freedom (DOF) $\omega$, \eqref{eq:sigmatotal}. Most DOF contribute to the positive part denoted as $\sigma_+$, however some of them contribute to the negative part $\sigma_-$. The total $\sigma=\sigma_+-\sigma_-$ is positive according to the second law of thermodynamics, \eqref{eq:sigmatotal}. \tcb{We call negentropy the negative contribution $\sigma_-$, and we hypothesize that $\sigma_-$, although much smaller in magnitude than $\sigma_+$, is the distinct signature of living versus inanimate matter.}

\begin{figure*}[t]
    \centering
    \includegraphics[width=0.6\textwidth]{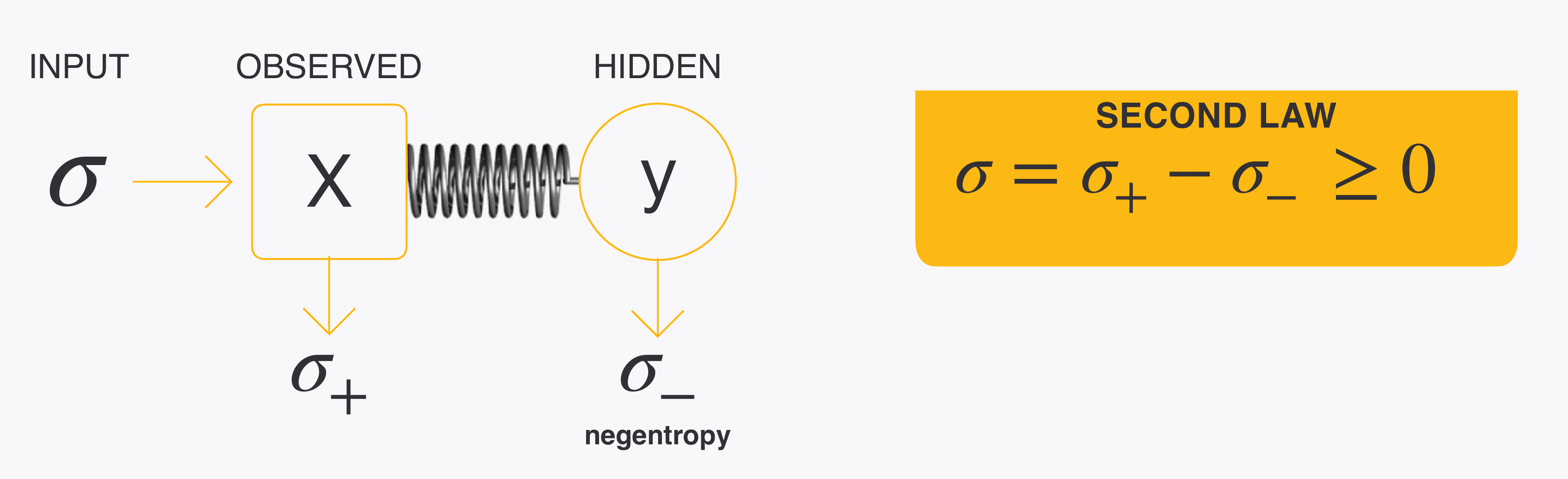}
    \caption{\textbf{The role of hidden DOFs.} Illustration of a minimal theoretical framework incorporating one observed and one hidden DOF. \tcb{The second law imposes $\sigma\ge 0$, but this does not preclude endothermic hidden DOFs with negentropy $\sigma_-$ that coexist with the large, positive exothermic contribution $\sigma_+$.} 
    }
    \label{fig:hidden}
\end{figure*}

In the living cell, the energy supplied by the sources, inorganic and organic matter, sunlight, etc., is diverted
into a main dissipative, exothermic and positive component $\sigma_+>0$ released as heat, plus an endothermic and negative
component, $-\sigma_-< 0$, which is used for remodeling and maintenance of cell functioning in a NESS. Endothermic negative contributions to $\sigma$ reflect the notion of negentropy
introduced by Schr\"odinger in his 1944 monograph \cite{schrodinger2025life} "What is Life?": \textit{"What an organism feeds upon is negative
entropy. Or, to put it less paradoxically, the essential thing in metabolism is that the organism succeeds in freeing
itself from all the entropy it cannot help producing while alive."}. Is it possible to experimentally measure the contributions $\sigma_+$ and $\sigma_-$? If yes, how? Recently, we have shown it is possible to extend the VSR to include \tcb{hidden degrees} of freedom following the treatment developed by Caldeira and Leggett for macroscopic quantum dissipation \cite{caldeira1983quantum}. \tcb{The approach follows the spirit of the reduced version of the Variance Sum Rule (reduced-VSR) discussed in Sec.\ref{subsec:model}, where one introduces hidden degrees of freedom in the model that allow reconstruction of the net force acting on the flickering probe (Fig. \ref{fig:hidden})}. This leads to a non-Markovian form of the VSR that can be directly applied to the experimental data \cite{DiTerlizziInPrep}. The total $\sigma$ can then be decomposed into the sum of the contributions of the observed DOF, \tcb{$\sigma_o$ (e.g., the position of the bead in the optical trap)}, and a hidden part $\sigma_h$ for the unobserved DOF modeled by another variable subject to active noise or non-reciprocal forces, $\sigma=\sigma_o+\sigma_h$. \tcb{We recently applied the method to derive the contributions $\sigma_o$ and $\sigma_h$ for an RNA molecule driven to a NESS by stochastically switching the pulling force via random telegraphic noise, implemented by moving the optical trap between two positions \cite{DiTerlizziInPrep}.} The observed DOF is the RNA molecular extension along the pulling axis, while the unobserved DOF is the  motion of the bead along the optical axial direction of the laser beam perpendincular to the pulling direction, relevant for both single- and dual-trap setups \cite{ribezzi2012force,ribezzi2015universal}. The resolution of the VSR-based approach permits to determine $\sigma_h$ contributions that are $\sim$10$\%$ of the total $\sigma$, of up to 0.1 attowatts or 100 zeptowatts (1 zW = $10^{-21}$ W; 1 $k_B T/s$). In this experimental realization, both contributions, $\sigma_o$ and $\sigma_h$, are positive, but nothing prevents a negative $\sigma_h$ in other situations. The VSR can reliably measure such tiny powers by combining high-time resolution flickering measurements with NESS modeling.  We stress the novelty of the VSR in measuring negentropy contributions, a feat out of reach for nanocalorimeters that \tcb{measure only} the total positive $\sigma$.

\subsection{\label{subsec:linearpol}Nucleic acids polymerization}
The applicability of the VSR to chemical reactions opens the possibility of investigating the energetics of metabolic processes with unprecedented accuracy. ATPases are suitable enzyme models to conduct such studies \cite{podolsky1956enthalpy,nicholls2013bioenergetics,ghosh2021enzymes}. They coordinate the hydrolysis of ATP in the presence of a substrate, co-factors and ions. Examples are translocases and polymerases that typically move along DNA at speeds $10-10^3$ base pairs per second (bp/s). As they typically hydrolyze a few ATPs per bp, $\sigma$ is bound by the available free energy from ATP hydrolysis $\Delta G\sim 10-15$kcal/mol $\sim 15-25$ $k_BT$ , giving $\sigma=10^2-10^4$ $k_BT/s$ in physiological conditions. Therefore, heat powers for ATP-dependent enzymes drop down to attoWatts (1 aW = $ 10^ {- 18} $ W, $\sim 10^3$ $k_BT/s$) or even tens of zeptowatts (1zW=$10^{-21}$ W, $\sim 1 k_BT/s$), a billion times ($10^9$) \tcb{beyond} the bar of nW in nanocalorimetry. 

A model of a NESS with negentropy production is the polymerization reaction, for example, during nucleic acid synthesis in DNA replication and transcription of DNA into RNA. According to Landauer, information is physical \cite{landauer1961irreversibility,bennett1982thermodynamics}: \tcb{its erasure increases the entropy of the environment by at least $ 1k_B\log2$ per bit}, the so-called Landauer \tcb{bound}. Therefore, the synthesis of new strands by DNA and RNA polymerases contributes to $\sigma_-$, while the total $\sigma$ during the polymerization reaction remains positive.

We can estimate the amount of negentropy rate in this case. The newborn nucleic acid strand is a unique sequence among all possible sequences that can be randomly generated one nucleotide at a time. The information content of the newborn strand grows by two bits per added nucleotide, corresponding to the four possible letters A,G,C,T(U), resulting in a total of 2n bits for a polymer of n nucleotides in length, assuming that concentrations and binding affinities of all nucleotides are equal. The synthesis of a new strand implies a decrease in entropy equal to $k_B \log 2$ per bit, and a negative $\sigma_-$ equal to $2nk_BT\log 2/t_n=2k_BT \log 2/t_1$, where $t_1=t_n/n $ is the average dwell time per nucleotide. Although reaction rates vary between DNA and RNA polymerases, they typically move at speeds $v\sim 100$ nucleotides per second (nt/s) and $t_1=1{\rm nt}/v\sim 0.01$s giving $\sigma_-=2k_BT\log 2/t_1\sim 100$ $k_BT/s\sim 0.4$ aW; therefore, negentropy rates for polymerases $\sigma_-$ are $\sim 10\%$ of the total $\sigma\sim 10^3$ $k_BT/s$ and within reach of the VSR.

A caveat is that polymerization can be a non-steady process, for example in single-molecule replication assays where the generation of the new strand changes the relative amount of double- versus single-stranded DNA. Because these two polymer forms have different mechanical properties, the overall stiffness of the system changes with time, and the replicative dynamics deviates from a NESS.  \tcb{However, that change is often mild under experimental conditions over the relevant timescales, and polymerization can be regarded as a NESS to a good degree.} Alternatively, one might extend the VSR approach to include a deterministic time-dependence in the model parameters to derive $\sigma$ in a quasi-NESS.
\begin{figure*}[t]
    \centering
    \includegraphics[width=0.6\textwidth]{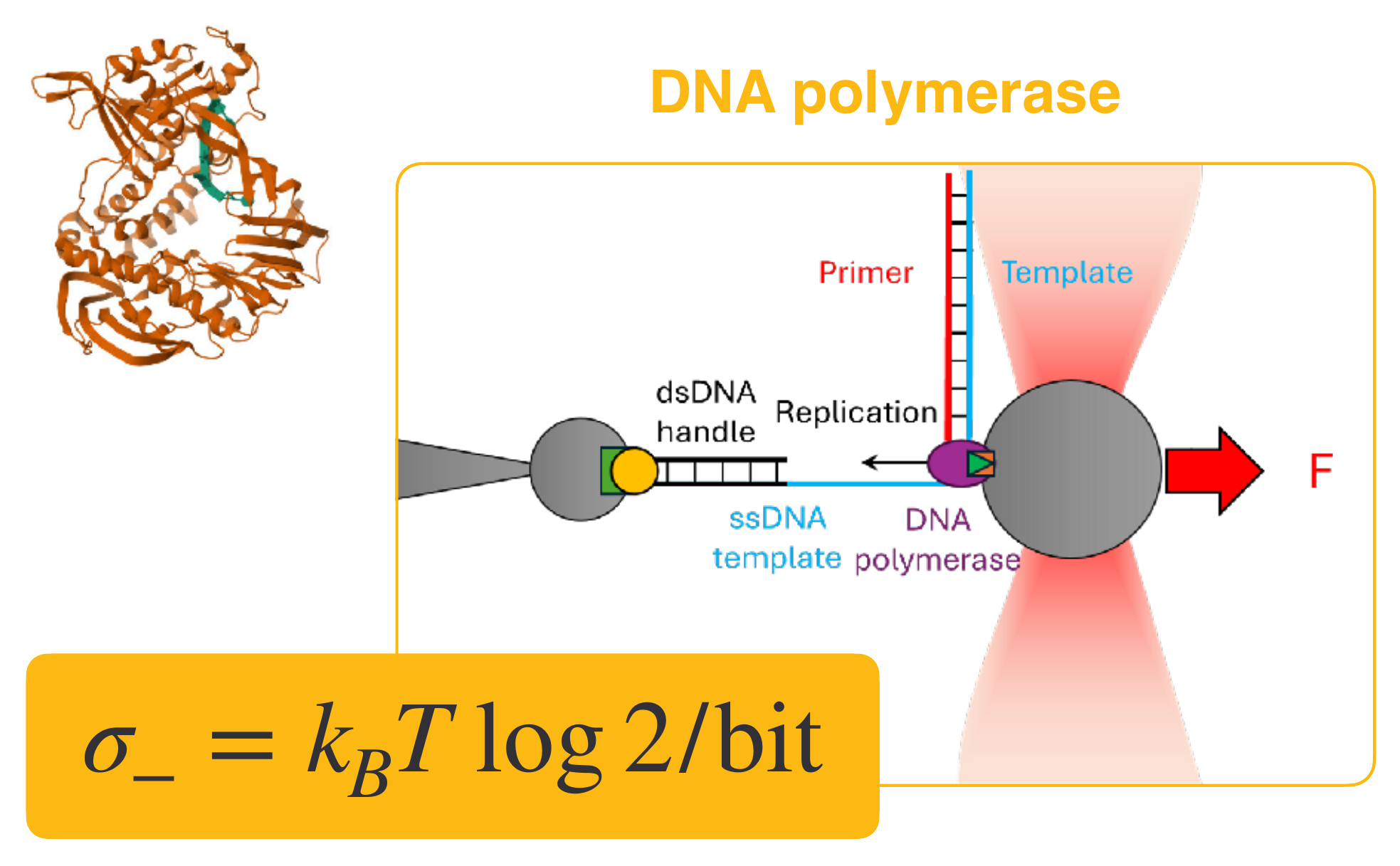}
    \caption{\textbf{Polymerization produces negentropy.} Illustration of an optical trapping assay where a DNA polymerase copies a parental ssDNA into a newborn strand, each nucleotide added contributes to negentropy with $k_BT\log 2$ per bit. At equal concentrations of the four nucleotides (A,G,C,T) this would amount to two bits per added nucleotide. 
    }
    \label{fig:polymerization}
\end{figure*}
\subsection{\label{subsec:linearheatflow}The linear heat flow example}
It is instructive to consider a somewhat trivial example where negentropy naturally emerges as an endothermic process unrelated to information content generation. \tcb{This is the case of steady heat flow between two reservoirs or sources, where the source receiving heat represents an exothermic DOF (it increases entropy) and the source delivering heat represents an endothermic DOF (it decreases entropy)}. The sources are kept at different temperatures, hot and cold, $T_H$ and $T_C$. Since no work is performed, energy conservation implies that the heat released by the hot source $-Q_H$ equals the heat absorbed by the cold source $Q_C$, or $Q_C=-Q_H\equiv Q$, with $Q$ a positive quantity. Correspondingly, the entropy of the hot source decreases by the amount $\Delta S_H=Q_H/T_H=-Q/T_H$ while that of the cold source increases by the amount $\Delta S_C=Q_C/T_C=Q/T_C$. Heat flow implies that the total entropy increases by $\Delta S=\Delta S_H+\Delta S_C=Q(1/T_C-1/T_H)=Q(T_H-T_C)/(T_HT_C)>0$. Therefore, the entropy production rate $\sigma=\Delta \dot{S}$ is directly proportional to $\dot{Q}$. In the linear regime where the Fourier law holds, we have $\dot{Q}=\sigma_0(T_H-T_C)$, with $\sigma_0$ a positive quantity directly proportional to the thermal conductivity between the reservoirs. Note that $\sigma_0$ has dimensions of entropy production rate ($k_B$ per unit time). From these results and defining $\sigma_H=\Delta \dot{S}_H$, $\sigma_C=\Delta \dot{S}_C$ we trivially obtain
\begin{equation}\label{eq:sigmaflow}
    \sigma=\Delta \dot{S}=\frac{\dot{Q}(T_H-T_C)}{T_HT_C}=\sigma_0\frac{(T_H-T_C)^2}{T_HT_C}\ge 0.
\end{equation}
with
\begin{eqnarray}\label{eq:sigmaflow2}
    \left\{
    \begin{array}{rcl}
    \sigma_C=\Delta \dot{S}_C
        &=& 
       \sigma_0\frac{(T_H-T_C)}{T_C}>0
       \\
        \sigma_H=\Delta \dot{S}_H
        &=& 
       -\sigma_0\frac{(T_H-T_C)}{T_H}<0
        \\
        \sigma
        &=& \sigma_C+\sigma_H 
    \end{array}
    \right.
\end{eqnarray}
Therefore, the hot source releases energy at a negentropy rate $\sigma_-=-\sigma_H$ that is lower than the rate at which positive entropy $\sigma_+=\sigma_C$ is produced at the cold source, $\sigma_H/\sigma_C=-T_C/T_H$ or $\sigma_-/\sigma_+=T_C/T_H$ according to \eqref{eq:sigmatotal}. 

\begin{figure*}[t]
    \centering
    \includegraphics[width=0.4\textwidth]{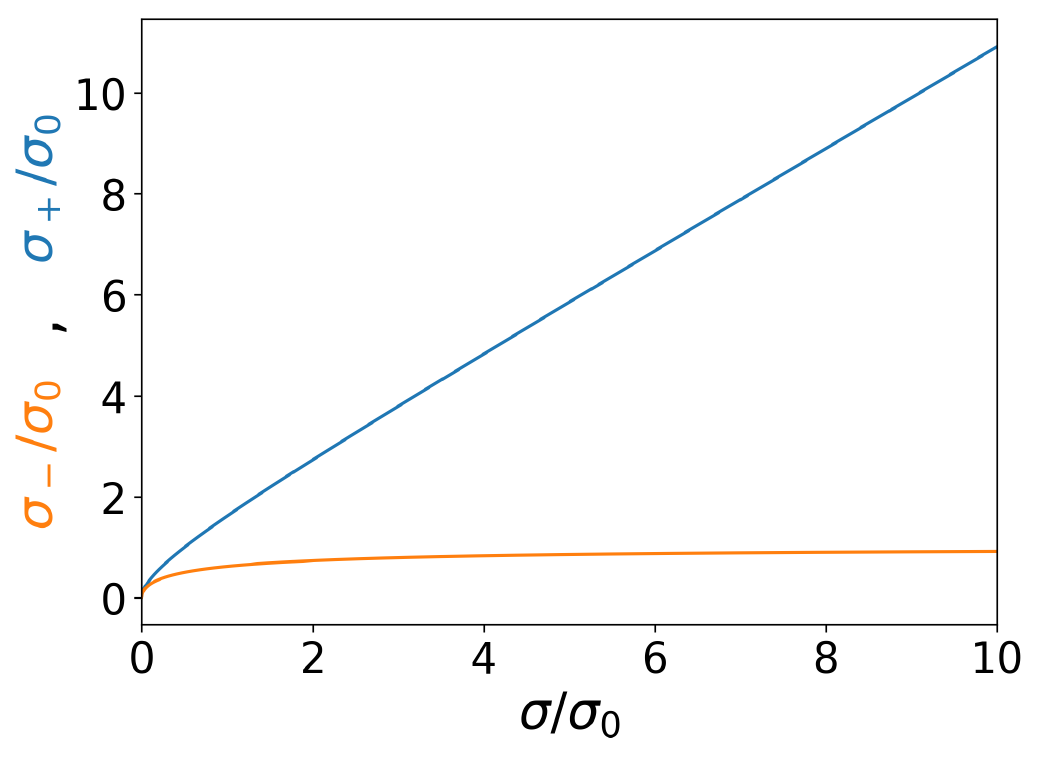}
    \caption{\textbf{The linear heat flow example.} Positive $\sigma_+/\sigma_0$ (blue) and negentropy $\sigma_-/\sigma_0$ (orange) contributions versus $\sigma/\sigma_0$, cf. \eqref{eq:sigmaflow3}. While $\sigma_+/\sigma_0$ grows with $\sigma/\sigma_0$, the negentropy part $\sigma_-/\sigma_0$ converges to 1, cf. \eqref{eq:sigmaflowlimits2}. This indicates that a finite negentropy $\sigma_-\sim\sigma_0$ requires a large $\sigma$. 
    }
    \label{fig:heatflow}
\end{figure*}

From \eqref{eq:sigmaflow2} we can express $\sigma_+\equiv\sigma_C$ and $\sigma_-\equiv-\sigma_H$ in terms of $\sigma$,
\begin{eqnarray}\label{eq:sigmaflow3}
    \left\{
    \begin{array}{rcl}
        
        \sigma_+\equiv\sigma_C&=&\sigma\frac{1+\sqrt{1+4\sigma_0/\sigma}}{2}
        \\
        \sigma_-\equiv-\sigma_H
        &=& \sigma\frac{-1+\sqrt{1+4\sigma_0/\sigma}}{2}
    \end{array}
    \right.
\end{eqnarray}
We consider the limits $\sigma\ll\sigma_0$ and $\sigma\gg\sigma_0$. By solving \eqref{eq:sigmaflow2} one trivially finds for $\sigma\ll\sigma_0$,
\begin{eqnarray}\label{eq:sigmaflowlimits}
    \left\{
    \begin{array}{rcl}
    \sigma_+&=&\sigma_0[\bigl(\frac{\sigma}{\sigma_0} \bigr)^{1/2}+\frac{\sigma}{2\sigma_0} +{\cal O}\bigl(\frac{\sigma}{\sigma_0} \bigr)^{3/2}]
        \\
        \sigma_-
        &=& \sigma_0[\bigl(\frac{\sigma}{\sigma_0} \bigr)^{1/2}-\frac{\sigma}{2\sigma_0} +{\cal O}\bigl(\frac{\sigma}{\sigma_0} \bigr)^{3/2}]
    \end{array}
    \right.
\end{eqnarray}
and both hot and cold source have negligible $\sigma_C(\equiv\sigma_+)\simeq -\sigma_H(\equiv \sigma_-)$. In contrast, for $\sigma\gg\sigma_0$,
\begin{eqnarray}\label{eq:sigmaflowlimits2}
   \left\{
    \begin{array}{rcl}
        
        \sigma_+&=&\sigma+\sigma_0[1+{\cal O}(\frac{\sigma_0}{\sigma})]
        \\
        \sigma_-&=&\sigma_0[1+{\cal O}(\frac{\sigma_0}{\sigma})]
    \end{array}
  \right.
\end{eqnarray}
so negentropy $\sigma_-=-\sigma_H\sim \sigma_0$ remains finite and determined by $\sigma_0$, while $\sigma_+=\sigma_C$ grows almost proportionally to the total $\sigma$. A plot of \eqref{eq:sigmaflow3} is shown in Figure \ref{fig:heatflow}. This result resonates with the fact that the large average heat power density $\sigma\gg \sigma_0$ is a necessary condition for a significant negentropy $\sigma_-$ ultimately determined by the thermal conductivity $\sigma_0$ and the concurrent heat flow, $\dot{Q}=\sigma_0(T_H-T_C)$.  Heat diffusion in the linear regime is too simple to capture any interesting behavior, and the limit $\sigma\gg\sigma_0$ might take the system beyond the linear regime of Fourier's law and linear irreversible thermodynamics. This example \tcb{shows} that negentropy can remain finite and much smaller than $\sigma$. Similar results are obtained with statistical models such as the Brownian gyrator model, a system of two mechanically coupled oscillators described by the DOF $x$ and $y$, in contact with two reservoirs at temperatures $T_x$ and $T_y$, where $\sigma$ was investigated using the VSR approach \cite{di2024variance2}.

\section{\label{sec:moluni}From molecules to the universe}
The average life heat power ${\cal P}_{\rm life}\simeq$1 W/kg is conserved across scales, from mammals to bacteria at the organismic level, down to individual cells and biomolecular reactions. \tcb{Variations exist depending on the system and environmental conditions, yet the range 0.1-10 W/kg encompasses most observed variability under stationary conditions.} The large heat dissipated by life contrasts with the comparatively low heat power density of the \tcb{Sun}, ${\cal P}_{\rm sun}\simeq 10^{-4}$ W/kg, and most stars in the universe, excluding non-stationary catastrophic events such as supernova explosions, gamma-ray bursts, etc. Among the highest emitting power density stars is $\eta \, \mathrm{Carinae}$, a binary system emitting about $10^{33}$ W, which has a mass of $\sim 10^{32}$ kg, giving 10 W/kg, comparable to average life power but still lower than the heat power density of a fibroblast or adipocyte, $\gtrsim 10-100$ W/kg. 

\begin{figure*}[t]
    \centering
    \includegraphics[width=0.8\textwidth]{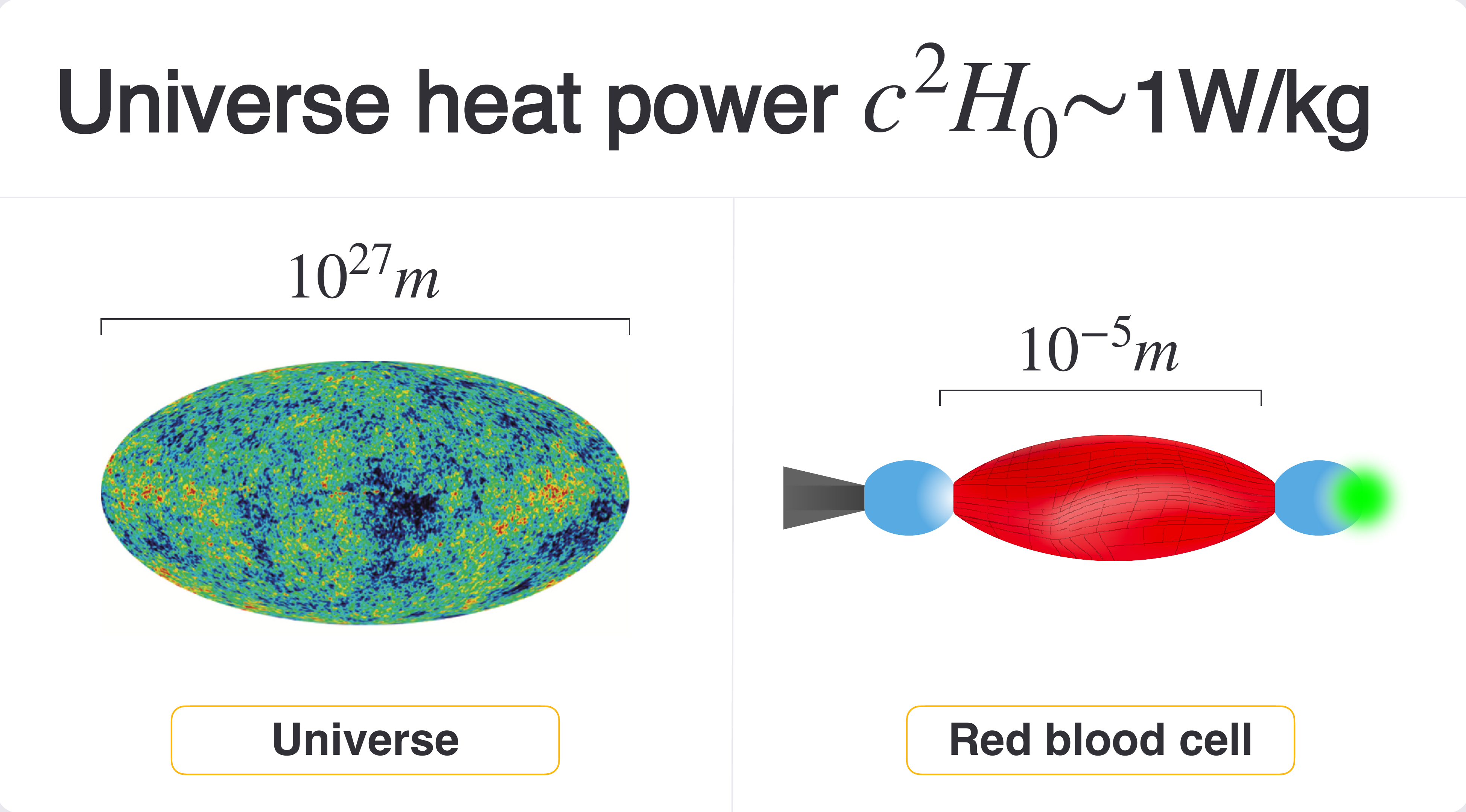}
    \caption{\textbf{Heat power density: from cells to the universe.} Despite the 32 orders of magnitude, the universe heat power falls in the range 1W/kg of living matter (here we show the cosmic microwave background and an RBC for illustrative purposes). 
    }
    \label{fig:heatuniverse}
\end{figure*}

\tcb{Remarkably,} ${\cal P}_{\rm life}\simeq$1W/kg is also the energy power density scale of the whole universe. This can be estimated as follows. A fundamental cosmological quantity with physical dimensions of power per unit mass is given by the product of the square of the speed of light $c$ divided by the universe age $t_\mathcal{U} \approx 13.8 \times 10^9 \,\mathrm{yr}$, approximately equal to the inverse of the Hubble constant $H_0 \approx 70 \,\mathrm{km\,s^{-1}\,Mpc^{-1}}$, giving the universe heat power density, 
\begin{equation}\label{eq:heatpoweruniverse}
    {\cal P}_{\mathcal{U}}=\frac{c^2}{t_\mathcal{U}}=c^2H_0=0.2\, {\rm W/kg}
\end{equation} 
which is comparable to ${\cal P}_{\rm life}\simeq$ 1W/kg. Note that the total power of the universe can also be expressed as the ratio of the energy content of the universe, $E_{0}=M_0 c^2$, divided by the universe age, $1/H_0$, giving a total power of $P_0=M_0 c^2H_0\sim 10^{54}$W with $M_0\sim 10^{54}$kg the current estimate for the mass of the universe, including ordinary and dark matter. \tcb{Therefore, we can also obtain the energy power density of the universe by dividing $P_0$ by $M_0$, so $M_0$ cancels in the ratio, giving \eqref{eq:heatpoweruniverse}}. Note that $P_0/M_0$ is the ratio of two large numbers $10^{54}$ W and $10^{54}$ kg. Its closeness to the power density of living matter \tcb{might} not be \tcb{a} coincidence (Fig.\ref{fig:heatuniverse}). 
Almost a century ago, Paul Dirac observed \cite{dirac1937cosmological} that the large dimensionless ratio of the electric and gravitational forces of electrons and protons in the atom $\sim 10^{39}-10^{40}$ must be related to the almost equally large universe age expressed in atomic time units $t_\mathcal{U} \approx 10^{34}t_0$, with $t_0$ the typical time of electron motion in atoms, roughly attoseconds or $10^{-18}$s. In Dirac's own words, \textit{The large dimensionless numbers appearing in physics are not accidental, but reflect a deep connection with the age of the universe.} In this way, Dirac inferred that the gravitational constant $G$ must decrease like $1/t$ with $t$ the age of the universe. Dirac's large number hypothesis has been disproved by experimental measurements of the rate of change of $G$, which are thousand times smaller than predicted by Dirac's theory, $\frac{\dot{G}}{G}\sim \frac{1}{t_\mathcal{U}}=10^{-10}$/year. Despite this, one might also envision that life's heat power decreases like $1/t_\mathcal{U}$ cf. \eqref{eq:heatpoweruniverse}, and that the value of ${\cal P}_{\rm life}$ was larger when life began to take its first steps than it is now and will be in the future with a variation rate $\frac{\dot{{\cal P}}_{\rm life}}{{\cal P}_{\rm life}}\sim \frac{1}{t_\mathcal{U}}=10^{-10}$/year. \tcb{Therefore, life may be a cosmological fact; it may have accompanied the expanding universe since eons ago in almost every corner.}   

It is instructive to estimate the heat power density ${\cal P}_{\rm CMB}$ of the cosmic microwave background radiation at $T_{\rm CMB}=2.7\,\mathrm{K}$, a proxy of ${\cal P}_{\mathcal{U}}$ in \eqref{eq:heatpoweruniverse}. The heat flux density of blackbody radiation follows the Stefan-Boltzmann law, $j_Q=\sigma T^4$, where $\sigma = 5.67 \times 10^{-8}\ \mathrm{W\,m^{-2}\,K^{-4}}$ is the Stefan-Boltzmann constant. The total heat power of the cosmic microwave background equals $P_{\rm CMB}=j_Q S_0\sim 10^{49}$W where $S_0=4\pi R_0^2$ is the so-called surface of last scattering and $R_0\sim 10^{27}$m the observable universe radius. 
A lower heat power density, ${\cal P}_{\rm CMB}=P_{\rm CMB}/M_0\sim 10^{-5}$W/kg is obtained, as expected, as the CMB radiation is only a fraction of ${\cal P}_{\mathcal{U}}$. It would be interesting to carry out such estimations with other widespread energy sources such as gravitational waves. Gravitational waves also exhibit flickering, and the power density could be estimated from stochastic gravity models \cite{hu2008stochastic}.

What is the role of negentropy in the global universe energy balance? We have seen in \eqref{eq:sigmatotal} that the total $\sigma$ can be decomposed as $\sigma=\sigma_+-\sigma_-$. Following \eqref{eq:heatpoweruniverse} one can envision a scenario of a young universe with a too large $\sigma$ for any significant negentropy $\sigma_-$ to trigger life start. \tcb{As cosmic acceleration drives the expanding universe, $\sigma$ decreases while $\sigma_-$ remains negligible until a threshold condition is reached; then $\sigma_-$ starts growing, and life spreads everywhere.} 
\begin{figure*}[t]
    \centering
    \includegraphics[width=0.6\textwidth]{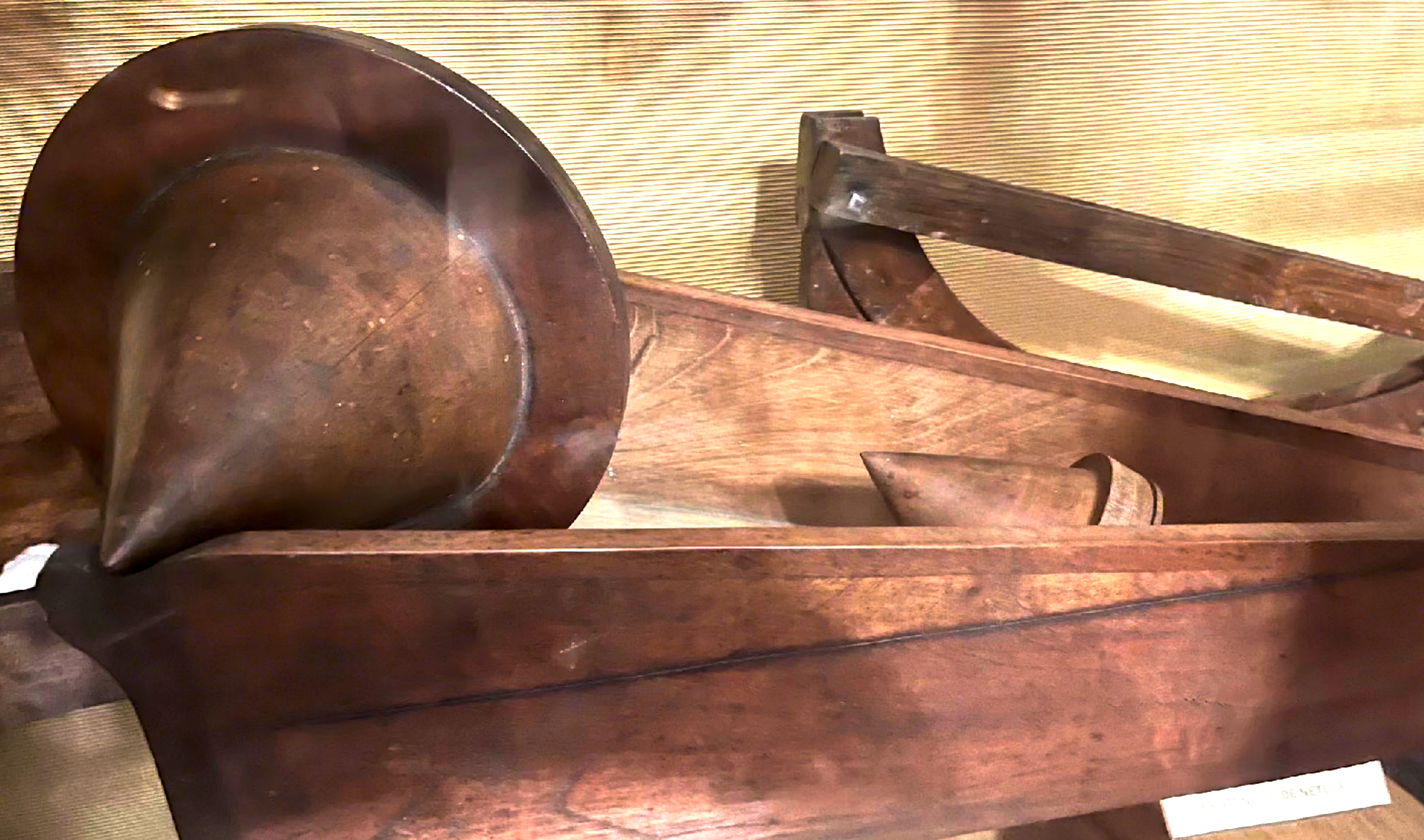}
    \caption{\textbf{Double cone paradox.} While the cone appears to move from right to left, rising along the edges of the wedge seemingly against gravity (analogous to negentropy), its center of mass is actually lowered (indicated by the dark line visible below the wedge), consistent with the action of gravity (analogous to $\sigma>0$).
    }
    \label{fig:doublecone}
\end{figure*}

An interesting analogy of a large but time-decreasing $\sigma$ and a small but time-increasing $\sigma_-$ can be made with the double cone mechanical paradox shown in Figure \ref{fig:doublecone}.) A double cone made out of two cones joined at their bases is mounted on diverging rails, and it appears to move uphill unexpectedly (corresponding to a finite $\sigma_->0$) while the center of mass is moving down as expected (corresponding to $\sigma>0$).  

\begin{figure*}[t]
    \centering
    \includegraphics[width=1.0\textwidth]{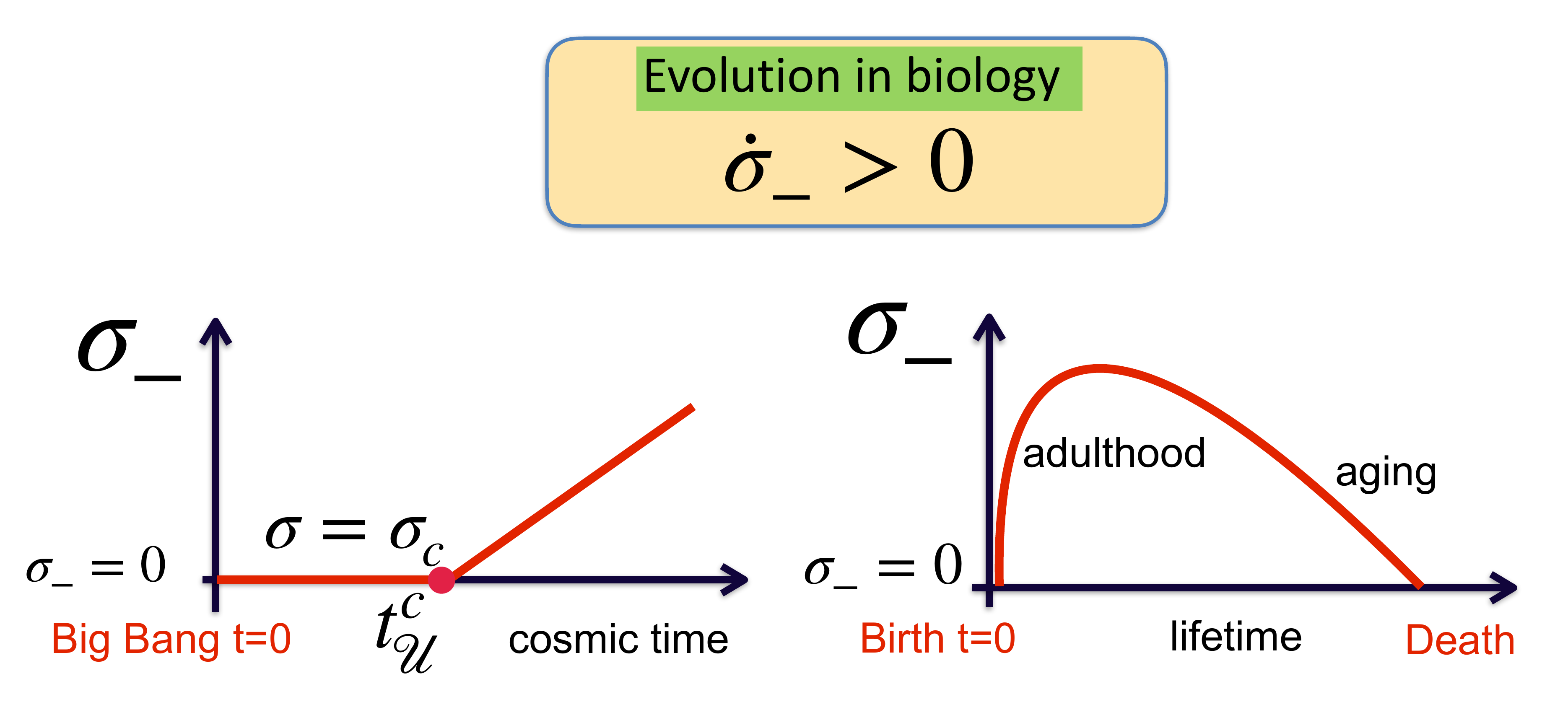}
    \caption{\textbf{Evolutionary scenarios.} Hypothetical negentropy in the universe driven by cosmic acceleration (left) and a living human across its life cycle (right). Biological evolution and growing complexity \tcb{are} characterized by an increasing negentropy over evolutionary timescales, $\dot{\sigma}_-> 0$. }
    \label{fig:evolution}
\end{figure*}

The decomposition of an overall positive $\sigma$ into a positive contribution ($\sigma_+$) and negentropy ($\sigma_-$) conforms to evolutionary biology where an increasing $\sigma_+$ on evolutionary timescales is accompanied by a corresponding increase in $\sigma_-$, while $\sigma$ remains constant or even decreases with cosmological time (Fig.\ref{fig:evolution}, left panel). Such a scenario is discussed below in Sec.\ref{sec:conclusions}. The balance $\sigma=\sigma_+-\sigma_-$ gives unbounded possibilities for natural selection \tcb{to make} negentropy grow in time. Any increase in $\sigma_-$ can be compensated by a corresponding increase in $\sigma_+$ without \tcb{affecting the overall $\sigma$, with the balance almost flirting with the second law.} Therefore, the distinguishing feature of the evolving tree of life might be stated as $\dot{\sigma}_->0$ (Fig.\ref{fig:evolution} top). One might speculate how negentropy in living matter shapes the life of organisms. \tcb{After conception and during development, an organism's negentropy increases rapidly, reaching a maximum at a given pre-adult stage, then declines again after adulthood during aging until it stops at death, where $\sigma=\sigma_+=\sigma_-=0$ (Fig.\ref{fig:evolution}, right panel).} 
\section{\label{sec:Discussion}Epilogue on the Variance Sum Rule (VSR)}
The VSR sets a dynamical constraint on the spectrum of fluctuations of the system typically over six decades of time. Together with the \textit{equation of state} for the passive and active fluctuations of the underlying model, the VSR permits \tcb{deriving} $\sigma$ in a model-independent way. This has \tcb{already been} tested for red blood cells (RBCs), \tcr{where simple models yield compatible values across different experimental setups.} The key is to incorporate the minimal number of degrees of freedom (DOF) necessary to capture the measured spectral density of the probe driven by the passive and active noise.

Is there always a VSR for a NESS? If the system exhibits active fluctuations, a proper model always sets a dynamical constraint and a VSR over several decades of time spanning the active timescales. Put differently, the VSR is the equivalent of the equation of state in thermodynamics but for the active fluctuations in a NESS. \tcr{A crucial requirement to apply the VSR and measure $\sigma$ is that the probe is coupled to the active DOF; otherwise, activity cannot be detected or distinguished from passive noise.} In mechanical measurements with \tcb{an} optically trapped bead as a probe, the stiffness must be lower than that of the system, a feasible goal in single molecule and cell measurements as already demonstrated for RBCs (Figs.\ref{fig:rigidity}, \ref{fig:RBC} and Ref. \cite{di2024variance}). Exothermic processes increase the fluctuations in the kinetic energy of molecules and ions inside the cell, inducing higher collision rates on the mechanical probe and active flickering. In this regard, the VSR approach might be applicable to investigate the enhanced diffusivity 
of chemically powered enzymatic nano- and micro-machines \cite{narlikar1997mechanistic,jee2018catalytic,xu2019direct,simmchen2024active,tang2025single}, a currently debated topic \cite{gunther2018diffusion,chen2020single,fillbrook2021following}.
\tcb{Similarly, FLIM measurements in cells are highly sensitive to electric fields from ions and protons in the cellular environment, which trigger fluctuations in the emission intensity, lifetime, and wavelength of fluorescent reporters.} Indeed, some studies have already reported large lifetime emission fluctuations of up to 100$\%$ in fluorophores embedded in polymer thin films \cite{vallee2003molecular}. Such large fluctuations have been observed in passive equilibrium conditions, which demonstrates the large sensitivity of emission lifetimes to the environment \cite{ma2024design}. In \tcb{the} presence of activity, one might expect fluctuations to be even larger. 

One may wonder whether heating effects due to the optical trap or laser excitation in FLIM do appreciably contribute to $\sigma$. \tcb{In general, a trapping laser in the mW range can produce a basal temperature increase of 0.1 Celsius over a micron-sized region, resulting in additional basal passive noise in the bead’s dynamics.} Such basal noise does not affect the value of $\sigma$ because the VSR incorporates the finite timescale of the active noise removing the effect of the passive noise from the analysis. Similarly, typical excitation powers in FLIM are on the order of $1\,\text{kW}/\text{cm}^2$. For the excitation wavelengths used for fluorophores in the visible $500-600$nm, the absorption fractions are small ($10^{-4}-10^{-6}$). As a result, the induced temperature increase is at most on the order of 0.1 degrees over a 1$\mu$m \tcb{lengthscale}, and even smaller over a 100nm region. This effect is therefore negligible, especially considering that nanothermometry studies report effective temperatures several degrees above the environmental temperature.

Are there other approaches to measure $\sigma$? Yes, but they fail for several reasons. Fluctuation theorems cannot measure $\sigma$ due to the inaccessibility of most configurational degrees of freedom and the limited statistics of forward-reverse trajectory pairs, while nanothermometry of cells only \tcb{measures} effective temperatures, ill-defined at the nanoscale, and proxies of the heat power at most (see above in Sec.\ref{sec:heatomics}). In contrast, the VSR requires monitoring nonequilibrium fluctuations of a few DOF at sufficiently high sampling rates to span the timescale of the cellular activity. The VSR approach has a \tcb{thermodynamic flavor}; it relies on the equation of state for the nonequilibrium fluctuations of the observed DOF. Finally, although nanocalorimetry is rapidly advancing toward sensitivities below the nanowatt scale \cite{lerchner2008nano,basta2018sensitive,hur2020sub,hong2020sub}, extending heat-power measurements to the subcellular or single-molecule level still remains out of reach.

While measuring the full heat power $\sigma$ appears feasible, it is still an open question how to measure negentropy. The connection between heat and information is rooted in the Landauer principle: the erasure of one bit of information \tcb{in the measurement device} dissipates heat \tcb{to the environment of} at least $k_BT\log 2$ per bit, making heat \tcb{a proxy for information}. \tcb{In polymerization reactions, negentropy might be measured by monitoring fluctuations in polymer extension at the single-molecule level and accounting for each added monomer at the Landauer rate of $k_BT\log 2$ per bit.} Which type of modeling is needed? While models for measuring $\sigma$ largely pertain to closed systems, for instance in RBCs and FLIM-based flickering measurements, negentropy $\sigma_-$ measurements might require \tcb{open-system models}, e.g., \tcb{models of polymerization} that exchange matter with a reservoir of nucleotides during DNA replication and RNA transcription.  

What type of NESS models are necessary to derive $\sigma$? It has been shown that models with a single DOF exhibiting Gaussian fluctuations cannot be used to discriminate nonequilibrium behavior and $\sigma$ \cite{bilotto2021excess,netz2025time}. In Ref.\cite{di2024variance}, the VSR was used to derive $\sigma$ in RBCs where fluctuations were Gaussian-like to a good \tcb{degree; however,} the applicability of the VSR is not restricted to Gaussian fluctuations, as we have seen for the case of the stochastic switching trap in Fig.\ref{fig:SST} and in non-linear quartic potential models (Sec. S14 in the Supp. Info. of Ref.\cite{di2024variance}). In this regard, molecular motor models described by kinetic rates for the polymerization reaction also exhibit non-Gaussian effects \cite{rodriguez2025continuous}. Strong non-Gaussian deviations are also expected in FLIM-flickering measurements, a challenge for models that use master equations with kinetic rates in an active environment \cite{pietzonka2024thermodynamic}. 

\tcb{In this vein,} we have emphasized the possibility of using the VSR approach to model FLIM-based flickering measurements. One could envision a kinetic model for light absorption and emission with three levels: the ground state and the excited singlet states, and the “inactive” or slowly decaying phosphorescent triplet state and other non-productive intersystem crossings that reduce the quantum yield. The VSR approach might treat the fluorescence emission signal between excited and ground states, also including fluctuating energy levels for the inter-crossing transitions and fluctuating quantum yields. Similar approaches based on fluctuating \tcb{kinetic rates} are also present in fluid turbulence, neuronal media, and other systems exhibiting intermittency effects. In these cases, transport coefficients and kinetic rates are treated as stochastic quantities. In this regard, the VSR approach might be applicable from passive turbulence to active fluids and neuroscience. 

In physics, an emerging field of study \tcb{is} active systems, and in particular active field theories \cite{nardini2017entropy,grandpre2021entropy,markovich2021thermodynamics} where the order parameter is the active $\sigma$-field across space and time, $\sigma(x,t)$, and the \tcb{analog} of a density, magnetization, or any order parameter field in statistical field theories. In Ref.\cite{di2024variance} we experimentally measured the $\sigma$-density field across the equatorial rim of RBCs, finding spatial heterogeneity with \tcb{a correlation length of $\sim 600$ nm, the first measurement of its kind}.  Heatomics may provide further experimental validations of field theories for active matter \cite{fodor2016far,fodor2018statistical}.

\section{\label{sec:conclusions}Conclusions}
We can summarize the main facts about $\sigma$. \tcb{First, energy is the most precious and limited resource for life, and the entropy production rate $\sigma$ is the unavoidable consequence of nature's irreversibility. Therefore,} energy consumption and heat power are tightly regulated in life. Second, $\sigma>0$ as dictated by the second law, permits the generation of negentropy $\sigma_-$ across hidden degrees of freedom (DOFs). Consequently, a higher $\sigma$ permits a larger negentropy cf. \eqref{eq:sigmatotal}, as reflected in the large average heat power per unit mass ${\cal P}_{\rm life}\simeq$ 1W/kg across life domains. Third, heat is the only communication signal that is naturally transported by diffusion and the second law. At the origin of life, heat diffusion across space was the only way for life entities to communicate before any \tcb{signaling molecules and signal-processing machinery existed. Fourth, when an organism dies, heat power stops, and $\sigma$ drops to 0. One could envision reverting such a transition} and restoring life again.  What is the energy cost for reverting such a transition? That would be the first step to \tcb{recover} animate from inanimate matter in the lab. 

Life is out of equilibrium and entropy production \tcb{is} its universal signature that makes the generation of negentropy possible, the part of the total heat power $\sigma$ that opposes degradation and death. The concept of negentropy was introduced by Schrödinger in 1944 in relation to information and maintenance of living matter. While it may look outdated, negentropy $\sigma_-$ is the essential part of life energetics that has never been measured, probably because it is a small fraction 0.1-10$\%$ of the total $\sigma$ in living matter, a tiny negative contribution distributed across many hidden DOFs and obscured by the large and positive contribution $\sigma_+$ such that $\sigma\simeq \sigma_+\gg \sigma_-$. \tcb{The Variance Sum Rule (VSR) is a theoretical framework that, combined with flickering data of a dynamical probe, permits experimentally deriving $\sigma$ with a resolution of attowatts ($ 10^ {- 18} $ W).} It can be generalized to incorporate multiple hidden degrees of freedom, paving the way to finding conditions in which negentropy can be produced and a living-NESS maintained.

Heatomics is the science of heat generation at the molecular and cellular level in living matter and the constraints imposed \tcb{on} biology. Why is heat power so high $\sim$ 1W/kg, and how is the small but essential negentropy part $\sigma_-$ generated and stably maintained? The idea that a heat-powered life requires chemical reaction networks with optimized parameters such as kinetic rates and energy and matter flows should be compatible with the existence of chemical cycles of finite negentropy \cite{schilling2025life}. Kinetic proofreading fidelity in genome maintenance and protein synthesis are examples that illustrate the importance of negentropy at the cost of large $\sigma$: each proofreading cycle dissipates heat, increasing $\sigma$. Thus, accuracy is paid for by dissipation, making $\sigma$ a direct measure of the thermodynamic cost of information processing and negentropy production.

The \tcb{study of nonequilibrium fluctuations to derive nanoscale energetics} goes hand in hand with the development of stochastic thermodynamics in nonequilibrium physics. This discipline has already permitted \tcb{the derivation of} free energy differences from irreversible work measurements. Examples are the work fluctuation theorem by Crooks and its corollary, the Jarzynski equality. In fact, work can be written as the product of generalized forces and displacements, whereas heat cannot. Work can be directly measured from the force exerted by a single external control parameter, e.g., using force spectroscopy tools, the trap position in optical tweezers, the cantilever height in an AFM, the position of the magnets in magnetic tweezers, and so on. \tcb{In contrast, heat is configurational energy associated with a large number of internal DOFs that are inaccessible and remain unobserved.} The traditional way to measure heat power is isothermal calorimetry in devices such as microfluidic chips or micromechanical systems that are kept in contact with the NESS while measuring the heat added or removed to maintain temperature's sample device constant. Current technology limits such measurements to nanowatts, a billion times higher than the attowatts typical of single molecular processes. In comparison, the resolution of 
the VSR down to attowatts opens many possibilities to quantify energy processes in biology.

In Sec.\eqref{sec:moluni} we underlined that the heat power in living matter ${\cal P}_{\rm life}\simeq$ 1 W/kg is ten thousand times \tcb{that of} the Sun. How is this possible? Despite its small size, the living cell produces heat everywhere at the sub-cellular level by slow combustion, from the nucleus to the endoplasmic reticulum, mitochondria, etc. In contrast, the Sun produces a large amount of heat non-uniformly, mainly at its core, a fraction of its size, taking an enormous amount of time for photons to reach the surface, about a hundred thousand years, making heat power per unit mass lower, ${\cal P}_{\rm sun}\simeq 10^{-4}$ W/kg. Not less important is that ${\cal P}_{\rm life}\simeq$ 1 W/kg is close to the average heat power density content of the entire universe ${\cal P}_{\cal U}\simeq$ 0.2 W/kg, the ratio of two big numbers, $\sim 10^{54}$, power (W) and mass (Kg), see Eq.\eqref{eq:heatpoweruniverse} in Sec.\eqref{sec:moluni}.

A correlation between the full heat power $\sigma$ and negentropy $\sigma_-$ might indicate optimized heat flows in living matter \cite{schilling2025life}. \tcb{A too-large $\sigma$ generates equally large heat flows, conditions unsuitable for stably sustaining a significant negentropy rate $\sigma_-$. Conversely, if $\sigma$ and heat flows are too small, the system is practically in equilibrium, and $\sigma_-$ is again residual.} An optimal NESS with a critical value $\sigma_c$ could induce internal correlations among the hidden DOF alongside with a significant negentropy stably maintained at the expense of $\sigma_c$. The value $\sigma_c$ could mark the onset of a dynamical transition separating two phases: $\sigma>\sigma_c$ where $\sigma_-$ is unstable, and life cannot be sustained; and $\sigma<\sigma_c$ where $\sigma_-$ can be stably kept, triggering a sustained negentropy growth over evolutionary timescales, $\dot{\sigma}_->0$. This is a speculative life scenario in an expanding universe that started with a large $\sigma={\cal P}_{\cal U}$ after the Big Bang \tcb{to subsequently decay while the universe cooled down}, cf. \eqref{eq:heatpoweruniverse} and Figure \ref{fig:evolution} (left panel). A critical $\sigma_c={\cal P}^c_{\cal U}$ after some critical universe age $t_{\cal U}^c$ determined the starting gun for the life race that began proliferating in some corners of the universe. \tcr{This moment could also coincide with the breaking of chiral symmetry \tcb{in prebiotic chemistry}, where molecular handedness symmetry was broken. The selection of D-sugars for nucleic acids and L-amino acids for proteins in living matter was the first negentropy event that reduced the entropy of the 50:50 L (laevus-left)-D (dexter-right) racemic abiotic mixtures.} \tcb{After $t_{\cal U}^c$, life has not only been stably maintained in the stable phase $\sigma={\cal P}_{\cal U}<\sigma_c={\cal P}^c_{\cal U}$, but it has also been evolving since then Figure \ref{fig:evolution}.}

The implications of heatomics go beyond biophysics and might find applications in other fields. Examples include complex systems such as active molecular condensates \cite{jawerth2020protein} and artificial motors \cite{hanggi2009artificial,colberg2014chemistry,kassem2017artificial,korosec2024motility}. In the field of condensed matter physics, one might envision applying the VSR to measure the heat power in macroscopic quantum dissipative regimes. Macroscopic degrees of freedom in systems such as Josephson junctions, nanomechanical resonators, Bose-Einstein condensates, nanomagnets, etc in a NESS exhibit flickering fluctuations that are informative of $\sigma$.
\begin{figure*}[t]
    \centering
    \includegraphics[width=0.6
\textwidth]{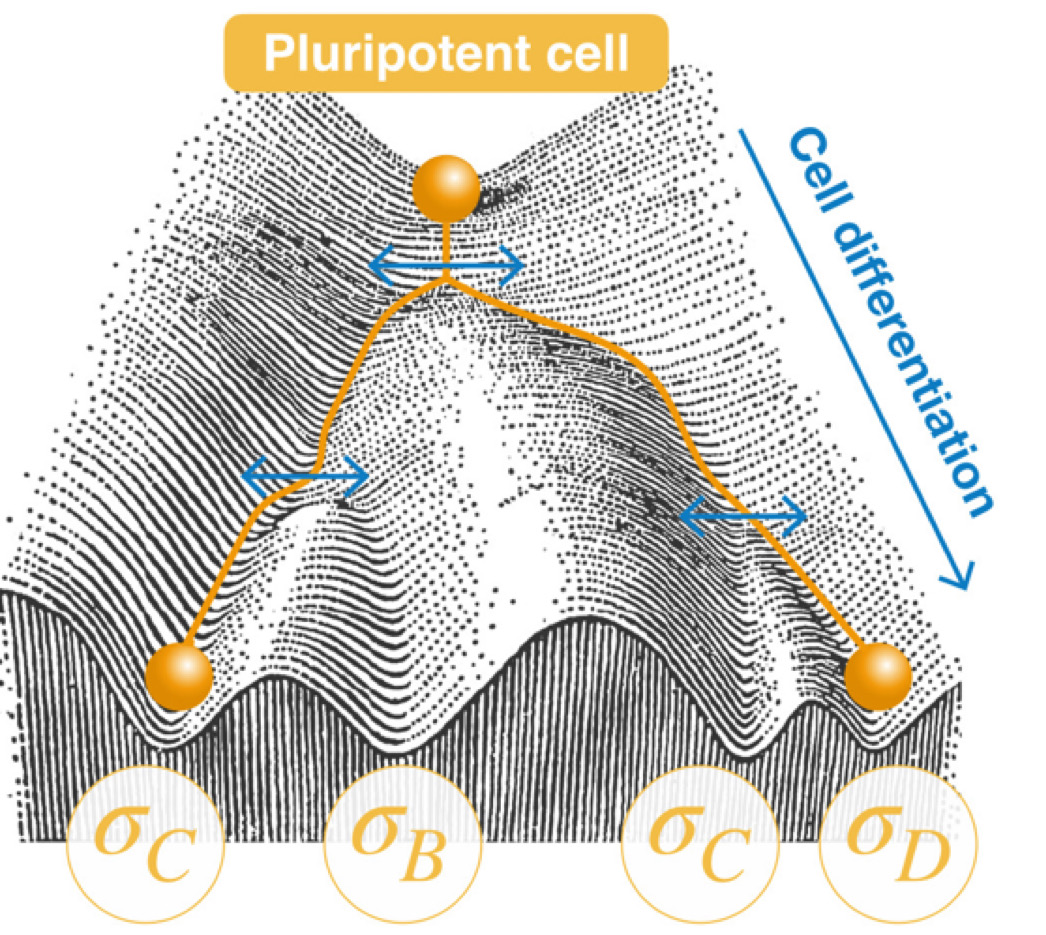}
    \caption{\textbf{$\sigma$-landscapes in biology.} Illustration of an evolutionary-developmental Waddington landscape where every cellular state is characterized by a different heat power $\sigma$.
    }
    \label{fig:development}
\end{figure*}

In cell-state theory in biology, cells transition between distinct stable states upon changing environmental cues \cite{mulas2021cell,rukhlenko2022control,rafelski2024establishing}. When this occurs $\sigma$ values change accordingly, suggesting of a new class of cellular $\sigma$-landscapes where each NESS corresponds to a minimum associated to a different $\sigma$ value, Figure \ref{fig:development}. The $\sigma$-landscapes are the NESS-equivalent of the free energy landscapes describing equilibrium conformations of molecular systems \cite{wang2008potential}. When a single cell changes state, its energetic $\sigma$-map also changes, maybe coherently across 
subcellular locations and scales, from the molecular to the cellular level. Besides $\sigma$-maps one might envision measuring negentropy $\sigma_-$-maps in living cells reflecting the energy and negentropy requirements at subcellular level. A paradigmatic analog of the $\sigma$-landscape is the Waddington landscape in developmental theory \cite{Huang2012}, where different developmental stages of a cell correspond to different phenotypes with their inherent $\sigma$ values (Figure \ref{fig:development}). Therefore, $\sigma$ appears as a unique descriptor of the NESS cellular state and a step towards quantifying energetically Waddington developmental landscapes.
\tcb{$\sigma$-landscapes open a paradigm for biological matter, enabling monitoring of how embryonic cells, cancer cells, and other pathological cell states change by correlating phenotypic and $\sigma$ changes.} Does $\sigma$ reflect
the functional heterogeneity and plasticity of cell states \cite{rafelski2024establishing} from normal to tumorigenic?  \tcb{Finally, let us add to this landscape the dead-equilibrium state characterized by $\sigma=0$, and envision perturbing a dead system to induce a transition to a non-zero finite $\sigma$, restoring the living state.} That would be thrilling. 

Finally, heatomics powered by the VSR approach might find applications to modern societal energy-related challenges, from climate change to energetic needs by AI. \tcr{At present, there is a strong need to reduce the energetic requirements of AI to approach the energetic efficiency of the brain. An example of forthcoming technologies for large-scale computations is neuromorphic ion computing in engineered ionic device architectures \cite{aluru2026knowledge}.  Among the many knowledge gaps, energetic efficiency stands out as a critical theme. Finally,} one might also envision measuring the local weather fluctuations, which together with VSR-based dynamical models of the atmosphere might permit monitoring climate's energy power and its evolution. Experimental measurements of $\sigma$ at all scales are most needed to illuminate the relevance of heat power for life. 
\section*{\label{sec:acknowledgments}Acknowledgments}
\tcr{I warmly thank Ivan Di Terlizzi, Marco Baiesi, Timo Betz, and Paco Monroy for a fruitful collaboration, Maria Garcia Parajo and Carlos Bustamante for stimulating discussions \tcb{on the relevance of heat power for cell-state transitions}, and all members of the Small Biosystems Lab for their continued support over the years. I thank Jort van Mourik, Antonio Grimaldi, and Eduard Vives for a critical reading of the manuscript.} I also acknowledge the Spanish Research Council Grant [PID2022-139913NB-I00] and the Generalitat de Catalunya through \tcb{the} ICREA Academia Prize 2023, for the financial support.
%
%
\bibliography{references}

@book{nicholls2013bioenergetics,
  title={Bioenergetics},
  author={Nicholls, David G},
  year={2013},
  publisher={Academic press}
}

@article{yang2021physical,
  title={Physical bioenergetics: Energy fluxes, budgets, and constraints in cells},
  author={Yang, Xingbo and Heinemann, Matthias and Howard, Jonathon and Huber, Greg and Iyer-Biswas, Srividya and Le Treut, Guillaume and Lynch, Michael and Montooth, Kristi L and Needleman, Daniel J and Pigolotti, Simone and others},
  journal={Proceedings of the National Academy of Sciences},
  volume={118},
  number={26},
  pages={e2026786118},
  year={2021},
  publisher={National Academy of Sciences}
}

@article{lan2012energy,
  title={The energy--speed--accuracy trade-off in sensory adaptation},
  author={Lan, Ganhui and Sartori, Pablo and Neumann, Silke and Sourjik, Victor and Tu, Yuhai},
  journal={Nature physics},
  volume={8},
  number={5},
  pages={422--428},
  year={2012},
  publisher={Nature Publishing Group UK London}
}

@article{cossetto2025thermodynamic,
  title={Thermodynamic dissipation constrains metabolic versatility of unicellular growth},
  author={Cossetto, Tommaso and Rodenfels, Jonathan and Sartori, Pablo},
  journal={Nature communications},
  volume={16},
  number={1},
  pages={8543},
  year={2025},
  publisher={Nature Publishing Group UK London}
}

@article{Makarieva2008,
  author = {Makarieva, Anastassia M. and Gorshkov, Victor G. and Li, Bai-Lian},
  title = {Mean mass-specific metabolic rates are strikingly similar across life's major domains: evidence for life's metabolic optimum},
  journal = {Proceedings of the National Academy of Sciences},
  volume = {105},
  number = {44},
  pages = {16994--16999},
  year = {2008},
  doi = {10.1073/pnas.0802148105},
  publisher = {National Academy of Sciences}
}

@article{Ballesteros2018,
  author = {Ballesteros, Fernando J. and Mart{\'i}nez, Vicente J. and Luque, Bartolo and Lacasa, Lucas and Valor, Enric},
  title = {On the Thermodynamic Origin of Metabolic Scaling},
  journal = {Scientific Reports},
  volume = {8},
  number = {1},
  pages = {1448},
  year = {2018},
  doi = {10.1038/s41598-018-19837-5},
  publisher = {Nature Publishing Group}
}

@book{Atkins2023,
  author = {Atkins, Peter William and De Paula, Julio and Keeler, James},
  title = {Atkins' Physical Chemistry},
  year = {2023},
  publisher = {Oxford University Press}
}

@book{Dill2010,
  author = {Dill, Ken and Bromberg, Sarina},
  title = {Molecular Driving Forces: Statistical Thermodynamics in Biology, Chemistry, Physics, and Nanoscience},
  year = {2010},
  publisher = {Garland Science}
}

@book{Kondepudi2015,
  author = {Kondepudi, Dilip and Prigogine, Ilya},
  title = {Modern Thermodynamics: From Heat Engines to Dissipative Structures},
  year = {2015},
  publisher = {John Wiley \& Sons}
}

@article{balasubramanian2021brain,
  title={Brain power},
  author={Balasubramanian, Vijay},
  journal={Proceedings of the National Academy of Sciences},
  volume={118},
  number={32},
  pages={e2107022118},
  year={2021},
  publisher={National Academy of Sciences}
}

@book{milo2015cell,
  title={Cell biology by the numbers},
  author={Milo, Ron and Phillips, Rob},
  year={2015},
  publisher={Garland Science}
}

@book{de2013non,
  title={Non-equilibrium thermodynamics},
  author={De Groot, Sybren Ruurds and Mazur, Peter},
  year={2013},
  publisher={Courier Corporation}
}

@article{maes2003origin,
  title={On the origin and the use of fluctuation relations for the entropy},
  author={Maes, Christian},
  journal={S{\'e}minaire Poincar{\'e}},
  volume={2},
  pages={29--62},
  year={2003}
}

@article{falasco2025macroscopic,
  title={Macroscopic stochastic thermodynamics},
  author={Falasco, Gianmaria and Esposito, Massimiliano},
  journal={Reviews of Modern Physics},
  volume={97},
  number={1},
  pages={015002},
  year={2025},
  publisher={APS}
}

@article{ritort2008nonequilibrium,
  title={Nonequilibrium fluctuations in small systems: From physics to biology},
  author={Ritort, Felix},
  journal={Advances in chemical physics},
  volume={137},
  pages={31},
  year={2008},
  publisher={Wiley Online Library}
}

@article{seifert2012stochastic,
  title={Stochastic thermodynamics, fluctuation theorems and molecular machines},
  author={Seifert, Udo},
  journal={Reports on progress in physics},
  volume={75},
  number={12},
  pages={126001},
  year={2012},
  publisher={IOP Publishing}
}

@article{ciliberto2017experiments,
  title={Experiments in stochastic thermodynamics: Short history and perspectives},
  author={Ciliberto, Sergio},
  journal={Physical Review X},
  volume={7},
  number={2},
  pages={021051},
  year={2017},
  publisher={APS}
}

@book{peliti2021stochastic,
  title={Stochastic thermodynamics: an introduction},
  author={Peliti, Luca and Pigolotti, Simone},
  year={2021},
  publisher={Princeton University Press}
}

@article{barato2015thermodynamic,
  title={Thermodynamic uncertainty relation for biomolecular processes},
  author={Barato, Andre C and Seifert, Udo},
  journal={Physical review letters},
  volume={114},
  number={15},
  pages={158101},
  year={2015},
  publisher={APS}
}

@article{horowitz2020thermodynamic,
  title={Thermodynamic uncertainty relations constrain non-equilibrium fluctuations},
  author={Horowitz, Jordan M and Gingrich, Todd R},
  journal={Nature Physics},
  volume={16},
  number={1},
  pages={15--20},
  year={2020},
  publisher={Nature Publishing Group UK London}
}

@article{dechant2021improving,
  title={Improving thermodynamic bounds using correlations},
  author={Dechant, Andreas and Sasa, Shin-ichi},
  journal={Physical Review X},
  volume={11},
  number={4},
  pages={041061},
  year={2021},
  publisher={APS}
}

@article{rodriguez2025continuous,
  title={Continuous-time random walk model for the diffusive motion of helicases},
  author={Rodr{\'\i}guez-Franco, Victor and Spiering, Michelle Marie and Bianco, Piero and Ritort, Felix and Manosas, Maria},
  journal={QRB discovery},
  volume={6},
  pages={e26},
  year={2025},
  publisher={Cambridge University Press}
}

@article{qian2004fluorescence,
  title={Fluorescence correlation spectroscopy with high-order and dual-color correlation to probe nonequilibrium steady states},
  author={Qian, Hong and Elson, Elliot L},
  journal={Proceedings of the National Academy of Sciences},
  volume={101},
  number={9},
  pages={2828--2833},
  year={2004},
  publisher={National Academy of Sciences}
}

@article{battle2016broken,
  title={Broken detailed balance at mesoscopic scales in active biological systems},
  author={Battle, Christopher and Broedersz, Chase P and Fakhri, Nikta and Geyer, Veikko F and Howard, Jonathon and Schmidt, Christoph F and MacKintosh, Fred C},
  journal={Science},
  volume={352},
  number={6285},
  pages={604--607},
  year={2016},
  publisher={American Association for the Advancement of Science}
}

@article{bisker2017hierarchical,
  title={Hierarchical bounds on entropy production inferred from partial information},
  author={Bisker, Gili and Polettini, Matteo and Gingrich, Todd R and Horowitz, Jordan M},
  journal={Journal of Statistical Mechanics: Theory and Experiment},
  volume={2017},
  number={9},
  pages={093210},
  year={2017},
  publisher={IOP Publishing and SISSA}
}

@article{teza2020exact,
  title={Exact coarse graining preserves entropy production out of equilibrium},
  author={Teza, Gianluca and Stella, Attilio L},
  journal={Physical Review Letters},
  volume={125},
  number={11},
  pages={110601},
  year={2020},
  publisher={APS}
}

@article{li2019quantifying,
  title={Quantifying dissipation using fluctuating currents},
  author={Li, Junang and Horowitz, Jordan M and Gingrich, Todd R and Fakhri, Nikta},
  journal={Nature communications},
  volume={10},
  number={1},
  pages={1666},
  year={2019},
  publisher={Nature Publishing Group UK London}
}

@article{manikandan2020inferring,
  title={Inferring entropy production from short experiments},
  author={Manikandan, Sreekanth K and Gupta, Deepak and Krishnamurthy, Supriya},
  journal={Physical review letters},
  volume={124},
  number={12},
  pages={120603},
  year={2020},
  publisher={APS}
}

@article{roldan2021quantifying,
  title={Quantifying entropy production in active fluctuations of the hair-cell bundle from time irreversibility and uncertainty relations},
  author={Rold{\'a}n, {\'E}dgar and Barral, J{\'e}r{\'e}mie and Martin, Pascal and Parrondo, Juan MR and J{\"u}licher, Frank},
  journal={New Journal of Physics},
  volume={23},
  number={8},
  pages={083013},
  year={2021},
  publisher={IOP Publishing}
}

@article{dieball2022mathematical,
  title={Mathematical, thermodynamical, and experimental necessity for coarse graining empirical densities and currents in continuous space},
  author={Dieball, Cai and Godec, Alja{\v{z}}},
  journal={Physical Review Letters},
  volume={129},
  number={14},
  pages={140601},
  year={2022},
  publisher={APS}
}

@article{pietzonka2024thermodynamic,
  title={Thermodynamic cost for precision of general counting observables},
  author={Pietzonka, Patrick and Coghi, Francesco},
  journal={Physical Review E},
  volume={109},
  number={6},
  pages={064128},
  year={2024},
  publisher={APS}
}

@article{parrondo2009entropy,
  title={Entropy production and the arrow of time},
  author={Parrondo, Juan MR and Broeck, C Van den and Kawai, Ryoichi},
  journal={New Journal of Physics},
  volume={11},
  number={7},
  pages={073008},
  year={2009}
}

@article{roldan2012entropy,
  title={Entropy production and Kullback-Leibler divergence between stationary trajectories of discrete systems},
  author={Rold{\'a}n, {\'E}dgar and Parrondo, Juan MR},
  journal={Physical Review E—Statistical, Nonlinear, and Soft Matter Physics},
  volume={85},
  number={3},
  pages={031129},
  year={2012},
  publisher={APS}
}

@article{skinner2021improved,
  title={Improved bounds on entropy production in living systems},
  author={Skinner, Dominic J and Dunkel, J{\"o}rn},
  journal={Proceedings of the National Academy of Sciences},
  volume={118},
  number={18},
  pages={e2024300118},
  year={2021},
  publisher={National Academy of Sciences}
}

@article{collin2005verification,
  title={Verification of the Crooks fluctuation theorem and recovery of RNA folding free energies},
  author={Collin, Delphine and Ritort, Felix and Jarzynski, Christopher and Smith, Steven B and Tinoco Jr, Ignacio and Bustamante, Carlos},
  journal={Nature},
  volume={437},
  number={7056},
  pages={231--234},
  year={2005},
  publisher={Nature Publishing Group UK London}
}

@article{junier2009recovery,
  title={Recovery of free energy branches in single molecule experiments},
  author={Junier, Ivan and Mossa, Alessandro and Manosas, Maria and Ritort, Felix},
  journal={Physical review letters},
  volume={102},
  number={7},
  pages={070602},
  year={2009},
  publisher={APS}
}

@article{alemany2012experimental,
  title={Experimental free-energy measurements of kinetic molecular states using fluctuation theorems},
  author={Alemany, Anna and Mossa, Alessandro and Junier, Ivan and Ritort, Felix},
  journal={Nature Physics},
  volume={8},
  number={9},
  pages={688--694},
  year={2012},
  publisher={Nature Publishing Group UK London}
}

@article{camunas2017experimental,
  title={Experimental measurement of binding energy, selectivity, and allostery using fluctuation theorems},
  author={Camunas-Soler, Joan and Alemany, Anna and Ritort, Felix},
  journal={Science},
  volume={355},
  number={6323},
  pages={412--415},
  year={2017},
  publisher={American Association for the Advancement of Science}
}

@article{rissone2022stem,
  title={Stem--loop formation drives RNA folding in mechanical unzipping experiments},
  author={Rissone, Paolo and Bizarro, Cristiano V and Ritort, Felix},
  journal={Proceedings of the National Academy of Sciences},
  volume={119},
  number={3},
  pages={e2025575119},
  year={2022},
  publisher={National Academy of Sciences}
}

@article{rico2022molten,
  title={Molten globule--like transition state of protein barnase measured with calorimetric force spectroscopy},
  author={Rico-Pasto, Marc and Zaltron, Annamaria and Davis, Sebastian J and Frutos, Silvia and Ritort, Felix},
  journal={Proceedings of the National Academy of Sciences},
  volume={119},
  number={11},
  pages={e2112382119},
  year={2022},
  publisher={National Academy of Sciences}
}

@article{ritort2002two,
  title={A two-state kinetic model for the unfolding of single molecules by mechanical force},
  author={Ritort, Felix and Bustamante, Carlos and Tinoco Jr, Ignacio},
  journal={Proceedings of the National Academy of Sciences},
  volume={99},
  number={21},
  pages={13544--13548},
  year={2002},
  publisher={National Academy of Sciences}
}

@article{gore2003bias,
  title={Bias and error in estimates of equilibrium free-energy differences from nonequilibrium measurements},
  author={Gore, Jeff and Ritort, Felix and Bustamante, Carlos},
  journal={Proceedings of the National Academy of Sciences},
  volume={100},
  number={22},
  pages={12564--12569},
  year={2003},
  publisher={National Academy of Sciences}
}

@article{lerchner2008nano,
  title={Nano-calorimetry of small-sized biological samples},
  author={Lerchner, J and Wolf, A and Schneider, H-J and Mertens, F and Kessler, E and Baier, V and Funfak, A and Nietzsch, M and Kr{\"u}gel, M},
  journal={Thermochimica acta},
  volume={477},
  number={1-2},
  pages={48--53},
  year={2008},
  publisher={Elsevier}
}

@article{basta2018sensitive,
  title={A sensitive calorimetric technique to study energy (heat) exchange at the nano-scale},
  author={Basta, Luca and Veronesi, Stefano and Murata, Yuya and Dubois, Zo{\'e} and Mishra, Neeraj and Fabbri, Filippo and Coletti, Camilla and Heun, Stefan},
  journal={Nanoscale},
  volume={10},
  number={21},
  pages={10079--10086},
  year={2018},
  publisher={Royal Society of Chemistry}
}

@article{hur2020sub,
  title={Sub-nanowatt resolution direct calorimetry for probing real-time metabolic activity of individual C. elegans worms},
  author={Hur, Sunghoon and Mittapally, Rohith and Yadlapalli, Swathi and Reddy, Pramod and Meyhofer, Edgar},
  journal={Nature communications},
  volume={11},
  number={1},
  pages={2983},
  year={2020},
  publisher={Nature Publishing Group UK London}
}

@article{hong2020sub,
  title={Sub-nanowatt microfluidic single-cell calorimetry},
  author={Hong, Sahngki and Dechaumphai, Edward and Green, Courtney R and Lal, Ratneshwar and Murphy, Anne N and Metallo, Christian M and Chen, Renkun},
  journal={Nature communications},
  volume={11},
  number={1},
  pages={2982},
  year={2020},
  publisher={Nature Publishing Group UK London}
}

@article{di2024variance,
  title={Variance sum rule for entropy production},
  author={Di Terlizzi, Ivan and Gironella, M and Herraez-Aguilar, D and Betz, Timo and Monroy, F and Baiesi, M and Ritort, F},
  journal={Science},
  volume={383},
  number={6686},
  pages={971--976},
  year={2024},
  publisher={American Association for the Advancement of Science}
}

@article{di2024variance2,
  title={Variance sum rule: proofs and solvable models},
  author={Di Terlizzi, Ivan and Baiesi, Marco and Ritort, Felix},
  journal={New Journal of Physics},
  volume={26},
  number={6},
  pages={063013},
  year={2024},
  publisher={IOP Publishing}
}

@article{roldan2024thermodynamic,
  title={Thermodynamic probes of life},
  author={Rold{\'a}n, {\'E}dgar},
  journal={Science},
  volume={383},
  number={6686},
  pages={952--953},
  year={2024},
  publisher={American Association for the Advancement of Science}
}

@article{noomnarm2009fluorescence,
  title={Fluorescence lifetimes: fundamentals and interpretations},
  author={Noomnarm, Ulai and Clegg, Robert M},
  journal={Photosynthesis research},
  volume={101},
  number={2},
  pages={181--194},
  year={2009},
  publisher={Springer}
}

@article{torrado2024fluorescence,
  title={Fluorescence lifetime imaging microscopy},
  author={Torrado, Belen and Pannunzio, Bruno and Malacrida, Leonel and Digman, Michelle A},
  journal={Nature Reviews Methods Primers},
  volume={4},
  number={1},
  pages={80},
  year={2024},
  publisher={Nature Publishing Group UK London}
}

@article{vallee2003molecular,
  title={Molecular fluorescence lifetime fluctuations: on the possible role of conformational effects},
  author={Vall{\'e}e, Renaud AL and Vancso, Gyula J and Van Hulst, NF and Calbert, J-P and Cornil, J and Bredas, JL},
  journal={Chemical physics letters},
  volume={372},
  number={1-2},
  pages={282--287},
  year={2003},
  publisher={Elsevier}
}

@article{fernandez2006master,
  title={A master relation defines the nonlinear viscoelasticity of single fibroblasts},
  author={Fern{\'a}ndez, Pablo and Pullarkat, Pramod A and Ott, Albrecht},
  journal={Biophysical journal},
  volume={90},
  number={10},
  pages={3796--3805},
  year={2006},
  publisher={Elsevier}
}

@article{puig2007viscoelasticity,
  title={Viscoelasticity of the human red blood cell},
  author={Puig-de-Morales-Marinkovic, Marina and Turner, Kevin T and Butler, James P and Fredberg, Jeffrey J and Suresh, Subra},
  journal={American Journal of Physiology-Cell Physiology},
  volume={293},
  number={2},
  pages={C597--C605},
  year={2007},
  publisher={American Physiological Society}
}

@article{gironella2024viscoelastic,
  title={Viscoelastic phenotyping of red blood cells},
  author={Gironella-Torrent, Marta and Bergamaschi, Giulia and Sorkin, Raya and Wuite, Gijs JL and Ritort, Felix},
  journal={Biophysical Journal},
  volume={123},
  number={7},
  pages={770--781},
  year={2024},
  publisher={Elsevier}
}

@article{zhou2025viscoelastic,
  title={Viscoelastic mechanics of living cells},
  author={Zhou, Hui and Liu, Ruye and Xu, Yizhou and Fan, Jierui and Liu, Xinyue and Chen, Longquan and Wei, Qiang},
  journal={Physics of Life Reviews},
  volume={53},
  pages={91--116},
  year={2025},
  publisher={Elsevier}
}

@article{backman1992microcalorimetric,
  title={A microcalorimetric study of human erythrocytes in stirred buffer suspensions},
  author={B{\"a}ckman, Per},
  journal={Thermochimica acta},
  volume={205},
  pages={87--97},
  year={1992},
  publisher={Elsevier}
}

@article{vetrone2010temperature,
  title={Temperature sensing using fluorescent nanothermometers},
  author={Vetrone, Fiorenzo and Naccache, Rafik and Zamarr{\'o}n, Alicia and Juarranz de la Fuente, Angeles and Sanz-Rodr{\'\i}guez, Francisco and Martinez Maestro, Laura and Martin Rodriguez, Emma and Jaque, Daniel and Garcia Sole, Jose and Capobianco, John A},
  journal={ACS nano},
  volume={4},
  number={6},
  pages={3254--3258},
  year={2010},
  publisher={ACS Publications}
}

@article{donner2012mapping,
  title={Mapping intracellular temperature using green fluorescent protein},
  author={Donner, Jon S and Thompson, Sebastian A and Kreuzer, Mark P and Baffou, Guillaume and Quidant, Romain},
  journal={Nano letters},
  volume={12},
  number={4},
  pages={2107--2111},
  year={2012},
  publisher={ACS Publications}
}

@article{okabe2012intracellular,
  title={Intracellular temperature mapping with a fluorescent polymeric thermometer and fluorescence lifetime imaging microscopy},
  author={Okabe, Kohki and Inada, Noriko and Gota, Chie and Harada, Yoshie and Funatsu, Takashi and Uchiyama, Seiichi},
  journal={Nature communications},
  volume={3},
  number={1},
  pages={705},
  year={2012},
  publisher={Nature Publishing Group UK London}
}

@article{kucsko2013nanometre,
  title={Nanometre-scale thermometry in a living cell},
  author={Kucsko, Georg and Maurer, Peter C and Yao, Norman Ying and Kubo, MICHAEL and Noh, Hyun Jong and Lo, Po Kam and Park, Hongkun and Lukin, Mikhail D},
  journal={Nature},
  volume={500},
  number={7460},
  pages={54--58},
  year={2013},
  publisher={Nature Publishing Group UK London}
}

@article{ebrahimi2014nucleic,
  title={Nucleic acid based fluorescent nanothermometers},
  author={Ebrahimi, Sara and Akhlaghi, Yousef and Kompany-Zareh, Mohsen and Rinnan, {\AA}smund},
  journal={Acs Nano},
  volume={8},
  number={10},
  pages={10372--10382},
  year={2014},
  publisher={ACS Publications}
}

@article{spicer2021harnessing,
  title={Harnessing DNA for nanothermometry},
  author={Spicer, Graham and Gutierrez-Erlandsson, Sylvia and Matesanz, Ruth and Bernard, Hugo and Adam, Alejandro P and Efeyan, Alejo and Thompson, Sebastian},
  journal={Journal of biophotonics},
  volume={14},
  number={2},
  pages={e202000341},
  year={2021},
  publisher={Wiley Online Library}
}

@article{chuma2024implication,
  title={Implication of thermal signaling in neuronal differentiation revealed by manipulation and measurement of intracellular temperature},
  author={Chuma, Shunsuke and Kiyosue, Kazuyuki and Akiyama, Taishu and Kinoshita, Masaki and Shimazaki, Yukiho and Uchiyama, Seiichi and Sotoma, Shingo and Okabe, Kohki and Harada, Yoshie},
  journal={Nature Communications},
  volume={15},
  number={1},
  pages={3473},
  year={2024},
  publisher={Nature Publishing Group UK London}
}

@article{baffou2014critique,
  title={A critique of methods for temperature imaging in single cells},
  author={Baffou, Guillaume and Rigneault, Herv{\'e} and Marguet, Didier and Jullien, Ludovic},
  journal={Nature methods},
  volume={11},
  number={9},
  pages={899--901},
  year={2014},
  publisher={Nature Publishing Group US New York}
}

@article{thompson2018plug,
  title={Plug and play anisotropy-based nanothermometers},
  author={Thompson, Sebastian A and Martinez, Ignacio A and Haro-Gonzalez, Patricia and P. Adam, Alejandro and Jaque, Daniel and Nieder, Jana B and de la Rica, Roberto},
  journal={ACS Photonics},
  volume={5},
  number={7},
  pages={2676--2681},
  year={2018},
  publisher={ACS Publications}
}

@article{zhou2020advances,
  title={Advances and challenges for fluorescence nanothermometry},
  author={Zhou, Jiajia and Del Rosal, Blanca and Jaque, Daniel and Uchiyama, Seiichi and Jin, Dayong},
  journal={Nature methods},
  volume={17},
  number={10},
  pages={967--980},
  year={2020},
  publisher={Nature Publishing Group US New York}
}

@article{chretien2018mitochondria,
  title={Mitochondria are physiologically maintained at close to 50 C},
  author={Chr{\'e}tien, Dominique and B{\'e}nit, Paule and Ha, Hyung-Ho and Keipert, Susanne and El-Khoury, Riyad and Chang, Young-Tae and Jastroch, Martin and Jacobs, Howard T and Rustin, Pierre and Rak, Malgorzata},
  journal={PLoS biology},
  volume={16},
  number={1},
  pages={e2003992},
  year={2018},
  publisher={Public Library of Science}
}

@book{schrodinger2025life,
  title={What is life? The physical aspect of the living cell},
  author={Schr{\"o}dinger, Erwin},
  year={2025},
  publisher={Rare Treasure Editions}
}

@article{caldeira1983quantum,
  title={Quantum tunnelling in a dissipative system},
  author={Caldeira, Amir O and Leggett, Anthony J},
  journal={Annals of physics},
  volume={149},
  number={2},
  pages={374--456},
  year={1983},
  publisher={Academic Press}
}

@article{ribezzi2012force,
  title={Force spectroscopy with dual-trap optical tweezers: Molecular stiffness measurements and coupled fluctuations analysis},
  author={Ribezzi-Crivellari, Marco and Ritort, Felix},
  journal={Biophysical Journal},
  volume={103},
  number={9},
  pages={1919--1928},
  year={2012},
  publisher={Elsevier}
}

@article{ribezzi2015universal,
  title={Universal axial fluctuations in optical tweezers},
  author={Ribezzi-Crivellari, Marco and Alemany, Anna and Ritort, Felix},
  journal={Optics Letters},
  volume={40},
  number={5},
  pages={800--803},
  year={2015},
  publisher={Optica Publishing Group}
}

@article{podolsky1956enthalpy,
  title={The enthalpy change of adenosine triphosphate hydrolysis},
  author={Podolsky, Richard J and Morales, Manuel F},
  journal={Journal of Biological Chemistry},
  volume={218},
  number={2},
  pages={945--959},
  year={1956},
  publisher={Elsevier}
}

@article{ghosh2021enzymes,
  title={Enzymes as active matter},
  author={Ghosh, Subhadip and Somasundar, Ambika and Sen, Ayusman},
  journal={Annual Review of Condensed Matter Physics},
  volume={12},
  number={1},
  pages={177--200},
  year={2021},
  publisher={Annual Reviews}
}

@article{landauer1961irreversibility,
  title={Irreversibility and heat generation in the computing process},
  author={Landauer, Rolf},
  journal={IBM journal of research and development},
  volume={5},
  number={3},
  pages={183--191},
  year={1961},
  publisher={Ibm}
}

@article{bennett1982thermodynamics,
  title={The thermodynamics of computation—a review},
  author={Bennett, Charles H},
  journal={International Journal of Theoretical Physics},
  volume={21},
  number={12},
  pages={905--940},
  year={1982},
  publisher={Springer}
}

@article{dirac1937cosmological,
  title={The cosmological constants},
  author={Dirac, Paul AM},
  journal={Nature},
  volume={139},
  number={3512},
  pages={323--323},
  year={1937},
  publisher={Nature Publishing Group UK London}
}

@article{hu2008stochastic,
  title={Stochastic gravity: Theory and applications},
  author={Hu, Bei Lok and Verdaguer, Enric},
  journal={Living Reviews in Relativity},
  volume={11},
  number={1},
  pages={1--112},
  year={2008},
  publisher={Springer}
}

@article{ma2024design,
  title={Design and application of fluorescent probes to detect cellular physical microenvironments},
  author={Ma, Junbao and Sun, Rui and Xia, Kaifu and Xia, Qiuxuan and Liu, Yu and Zhang, Xin},
  journal={Chemical reviews},
  volume={124},
  number={4},
  pages={1738--1861},
  year={2024},
  publisher={ACS Publications}
}

@article{bilotto2021excess,
  title={Excess and loss of entropy production for different levels of coarse graining},
  author={Bilotto, Pierpaolo and Caprini, Lorenzo and Vulpiani, Angelo},
  journal={Physical Review E},
  volume={104},
  number={2},
  pages={024140},
  year={2021},
  publisher={APS}
}

@article{netz2025time,
  title={Time-dependent trajectory of a one-dimensional Gaussian non-Markovian observable does not reveal its nonequilibrium character},
  author={Netz, Roland R},
  journal={Physical Review E},
  volume={112},
  number={1},
  pages={014132},
  year={2025},
  publisher={APS}
}

@article{nardini2017entropy,
  title={Entropy production in field theories without time-reversal symmetry: quantifying the non-equilibrium character of active matter},
  author={Nardini, Cesare and Fodor, {\'E}tienne and Tjhung, Elsen and Van Wijland, Fr{\'e}d{\'e}ric and Tailleur, Julien and Cates, Michael E},
  journal={Physical Review X},
  volume={7},
  number={2},
  pages={021007},
  year={2017},
  publisher={APS}
}

@article{grandpre2021entropy,
  title={Entropy production fluctuations encode collective behavior in active matter},
  author={GrandPre, Trevor and Klymko, Katherine and Mandadapu, Kranthi K and Limmer, David T},
  journal={Physical Review E},
  volume={103},
  number={1},
  pages={012613},
  year={2021},
  publisher={APS}
}

@article{markovich2021thermodynamics,
  title={Thermodynamics of active field theories: Energetic cost of coupling to reservoirs},
  author={Markovich, Tomer and Fodor, {\'E}tienne and Tjhung, Elsen and Cates, Michael E},
  journal={Physical Review X},
  volume={11},
  number={2},
  pages={021057},
  year={2021},
  publisher={APS}
}

@article{fodor2016far,
  title={How far from equilibrium is active matter?},
  author={Fodor, {\'E}tienne and Nardini, Cesare and Cates, Michael E and Tailleur, Julien and Visco, Paolo and Van Wijland, Fr{\'e}d{\'e}ric},
  journal={Physical review letters},
  volume={117},
  number={3},
  pages={038103},
  year={2016},
  publisher={APS}
}

@article{fodor2018statistical,
  title={The statistical physics of active matter: From self-catalytic colloids to living cells},
  author={Fodor, {\'E}tienne and Marchetti, M Cristina},
  journal={Physica A: Statistical Mechanics and its Applications},
  volume={504},
  pages={106--120},
  year={2018},
  publisher={Elsevier}
}

@article{schilling2025life,
  title={Why life is hot},
  author={Schilling, Tanja and Warren, Patrick B and Poon, Wilson},
  journal={arXiv preprint arXiv:2512.04725},
  year={2025}
}

@article{mulas2021cell,
  title={Cell state transitions: definitions and challenges},
  author={Mulas, Carla and Chaigne, Agathe and Smith, Austin and Chalut, Kevin J},
  journal={Development},
  volume={148},
  number={20},
  pages={dev199950},
  year={2021},
  publisher={The Company of Biologists Ltd}
}

@article{rukhlenko2022control,
  title={Control of cell state transitions},
  author={Rukhlenko, Oleksii S and Halasz, Melinda and Rauch, Nora and Zhernovkov, Vadim and Prince, Thomas and Wynne, Kieran and Maher, Stephanie and Kashdan, Eugene and MacLeod, Kenneth and Carragher, Neil O and others},
  journal={Nature},
  volume={609},
  number={7929},
  pages={975--985},
  year={2022},
  publisher={Nature Publishing Group UK London}
}

@article{rafelski2024establishing,
  title={Establishing a conceptual framework for holistic cell states and state transitions},
  author={Rafelski, Susanne M and Theriot, Julie A},
  journal={Cell},
  volume={187},
  number={11},
  pages={2633--2651},
  year={2024},
  publisher={Elsevier}
}

@article{wang2008potential,
  title={Potential landscape and flux framework of nonequilibrium networks: robustness, dissipation, and coherence of biochemical oscillations},
  author={Wang, Jin and Xu, Li and Wang, Erkang},
  journal={Proceedings of the National Academy of Sciences},
  volume={105},
  number={34},
  pages={12271--12276},
  year={2008},
  publisher={National Academy of Sciences}
}

@article{Huang2012,
  author = {Huang, Sui},
  title = {The molecular and mathematical basis of Waddington's epigenetic landscape: A framework for post-Darwinian biology?},
  journal = {BioEssays},
  volume = {34},
  number = {2},
  pages = {149--157},
  year = {2012},
  doi = {10.1002/bies.201100031}
}

@article{jawerth2020protein,
  title={Protein condensates as aging Maxwell fluids},
  author={Jawerth, Louise and Fischer-Friedrich, Elisabeth and Saha, Suropriya and Wang, Jie and Franzmann, Titus and Zhang, Xiaojie and Sachweh, Jenny and Ruer, Martine and Ijavi, Mahdiye and Saha, Shambaditya and others},
  journal={Science},
  volume={370},
  number={6522},
  pages={1317--1323},
  year={2020},
  publisher={American Association for the Advancement of Science}
}

@article{hanggi2009artificial,
  title={Artificial Brownian motors: Controlling transport on the nanoscale},
  author={H{\"a}nggi, Peter and Marchesoni, Fabio},
  journal={Reviews of Modern Physics},
  volume={81},
  number={1},
  pages={387--442},
  year={2009},
  publisher={APS}
}

@article{colberg2014chemistry,
  title={Chemistry in motion: tiny synthetic motors},
  author={Colberg, Peter H and Reigh, Shang Yik and Robertson, Bryan and Kapral, Raymond},
  journal={Accounts of chemical research},
  volume={47},
  number={12},
  pages={3504--3511},
  year={2014},
  publisher={ACS Publications}
}

@article{kassem2017artificial,
  title={Artificial molecular motors},
  author={Kassem, Salma and van Leeuwen, Thomas and Lubbe, Anouk S and Wilson, Miriam R and Feringa, Ben L and Leigh, David A},
  journal={Chemical Society Reviews},
  volume={46},
  number={9},
  pages={2592--2621},
  year={2017},
  publisher={Royal Society of Chemistry}
}

@article{korosec2024motility,
  title={Motility of an autonomous protein-based artificial motor that operates via a burnt-bridge principle},
  author={Korosec, Chapin S and Unksov, Ivan N and Surendiran, Pradheebha and Lyttleton, Roman and Curmi, Paul MG and Angstmann, Christopher N and Eichhorn, Ralf and Linke, Heiner and Forde, Nancy R},
  journal={Nature Communications},
  volume={15},
  number={1},
  pages={1511},
  year={2024},
  publisher={Nature Publishing Group UK London}
}

@misc{DiTerlizziInPrep,
  author = {Di Terlizzi, I. and others},
  note = {In preparation},
  year = {2026}
}

@article{narlikar1997mechanistic,
  title={Mechanistic aspects of enzymatic catalysis: lessons from comparison of RNA and protein enzymes},
  author={Narlikar, Geeta J and Herschlag, Daniel},
  journal={Annual review of biochemistry},
  volume={66},
  number={1},
  pages={19--59},
  year={1997},
  publisher={Annual Reviews 4139 El Camino Way, PO Box 10139, Palo Alto, CA 94303-0139, USA}
}

@article{xu2019direct,
  title={Direct single molecule imaging of enhanced enzyme diffusion},
  author={Xu, Mengqi and Ross, Jennifer L and Valdez, Lyanne and Sen, Aysuman},
  journal={Physical review letters},
  volume={123},
  number={12},
  pages={128101},
  year={2019},
  publisher={APS}
}

@article{jee2018catalytic,
  title={Catalytic enzymes are active matter},
  author={Jee, Ah-Young and Cho, Yoon-Kyoung and Granick, Steve and Tlusty, Tsvi},
  journal={Proceedings of the National Academy of Sciences},
  volume={115},
  number={46},
  pages={E10812--E10821},
  year={2018},
  publisher={National Academy of Sciences}
}

@article{tang2025single,
  title={Single molecule--driven nanomotors reveal the dynamic-disordered chemomechanical transduction of active enzymes},
  author={Tang, Zhuodong and Wu, Jingyu and Wu, Shaojun and Tang, Wenjing and Zhang, Jian-Rong and Zhu, Wenlei and Zhu, Jun-Jie and Chen, Zixuan},
  journal={Science Advances},
  volume={11},
  number={5},
  pages={eads0446},
  year={2025},
  publisher={American Association for the Advancement of Science}
}

@book{simmchen2024active,
  title={Active Colloids: From Fundamentals to Frontiers},
  author={Simmchen, Juliane and Uspal, William and Wang, Wei},
  year={2024},
  publisher={Royal Society of Chemistry}
}

@article{gunther2018diffusion,
  title={Diffusion measurements of swimming enzymes with fluorescence correlation spectroscopy},
  author={Gunther, Jan-Philipp and Borsch, Michael and Fischer, Peer},
  journal={Accounts of chemical research},
  volume={51},
  number={9},
  pages={1911--1920},
  year={2018},
  publisher={ACS Publications}
}

@article{chen2020single,
  title={Single-molecule diffusometry reveals no catalysis-induced diffusion enhancement of alkaline phosphatase as proposed by FCS experiments},
  author={Chen, Zhijie and Shaw, Alan and Wilson, Hugh and Woringer, Maxime and Darzacq, Xavier and Marqusee, Susan and Wang, Quan and Bustamante, Carlos},
  journal={Proceedings of the National Academy of Sciences},
  volume={117},
  number={35},
  pages={21328--21335},
  year={2020},
  publisher={National Academy of Sciences}
}

@article{fillbrook2021following,
  title={Following molecular mobility during chemical reactions: no evidence for active propulsion},
  author={Fillbrook, Lucy L and Gunther, Jan-Philipp and Majer, Gunter and O’Leary, Daniel J and Price, William S and Van Ryswyk, Hal and Fischer, Peer and Beves, Jonathon E},
  journal={Journal of the American Chemical Society},
  volume={143},
  number={49},
  pages={20884--20890},
  year={2021},
  publisher={ACS Publications}
}

@book{Cox2025Lehninger,
  author    = {Michael M. Cox and Aaron A. Hoskins and Alain Viel and Judith Simcox},
  title     = {Lehninger Biochemistry: Core Concepts and Applications},
  edition   = {1},
  year      = {2025},
  publisher = {W. H. Freeman},
  address   = {New York}
}

@article{aluru2026knowledge,
  title={Knowledge gaps for neuromorphic ionic computing},
  author={Aluru, Narayana R and Darling, Seth B and Elam, Jeffrey W and Gang, Oleg and Salleo, Alberto and Siwy, Zuzanna and Talin, A Alec and Noy, Aleksandr},
  journal={Science},
  volume={392},
  number={6798},
  pages={592--601},
  year={2026},
  publisher={American Association for the Advancement of Science}
}

@article{palassini2011improving,
  title={Improving free-energy estimates from unidirectional work measurements: theory and experiment},
  author={Palassini, Matteo and Ritort, Felix},
  journal={Physical Review Letters},
  volume={107},
  number={6},
  pages={060601},
  year={2011},
  publisher={APS}
}

@article{lee2026estimating,
  title={Estimating Free Energy Differences with Virtually Escorted Trajectories},
  author={Lee, Sangyun and Jarzynski, Christopher},
  journal={arXiv preprint arXiv:2606.30451},
  year={2026}
}
\end{document}